\documentclass[adp,fleqn]{w-art}
\usepackage{times}
\numberwithin{equation}{section}
\numberwithin{figure}{section}
\usepackage[]{graphicx}

\begin{document}
\DOIsuffix{theDOIsuffix}
\Volume{12}
\Issue{1}
\Copyrightissue{01}
\Month{01}
\Year{2003}
\pagespan{3}{}



\title[The first second of the Universe]{The first second of the Universe}

\author[D.~J.~Schwarz]{Dominik J.~Schwarz\footnote{
                                      e-mail:{\sf dominik.schwarz@cern.ch}} 
     \address{Theory Division, CERN, 1211 Geneva 23, Switzerland}} 

\begin{abstract}
The history of the Universe after its first second is now tested by high 
quality observations of light element abundances and temperature anisotropies 
of the cosmic microwave background. The epoch of the first second itself has 
not been tested directly yet; however, it is constrained by experiments at 
particle and heavy ion accelerators. Here I attempt to describe the  
epoch between the electroweak transition and the primordial nucleosynthesis. 

The most dramatic event in that era is the quark--hadron transition at 
$10$~$\mu$s. Quarks and gluons condense to form a gas of nucleons and 
light mesons, the latter decay subsequently. At the end of the first 
second, neutrinos and neutrons decouple from the radiation fluid. 
The quark--hadron transition and dissipative processes during the first 
second prepare the initial conditions for the synthesis of the first nuclei.

As for the cold dark matter (CDM), WIMPs (weakly interacting massive 
particles) --- the most popular candidates for the CDM --- decouple from the 
presently known forms of matter, chemically (freeze-out) at $10$~ns and 
kinetically at $1$~ms. The chemical decoupling fixes their present abundances
and dissipative processes during and after thermal decoupling set the 
scale for the very first WIMP clouds. 
\end{abstract}
\maketitle

\tableofcontents

\section{Introduction}

This article is an attempt to summarize our present understanding of the early 
Universe in the epoch between the electroweak (EW) transition and the onset
of primordial nucleosynthesis. During that time the temperature drops from  
$T_{\rm EW} \approx 100$--$200$ GeV at the EW transition to $1$~MeV, 
when all weak interaction rates fall below the Hubble rate, which marks the 
beginning of the primordial nucleosynthesis epoch. At $1$ MeV, the Hubble age, 
$t_{\rm H} \equiv 1/H$, is $1$~s; in that sense this is a review about 
the first second of the Universe.  The physics of the first second is 
well tested at high energy colliders, as far as the particle content up to 
masses of $\sim 100$ GeV is concerned. The equation of state of the radiation
fluid that dominates the Universe during the first second is not known up 
to the highest temperatures, but it is tested currently by heavy ion 
experiments up to the scale of quantum chromodynamics (QCD) at 
$100$--$200$~MeV. However, I would like to stress that, since details of 
the QCD transition and the nature of cold dark matter (CDM) are still unknown,
we have to rely on yet untested models. 

The Universe was radiation-dominated during the first second. Today, this 
conclusion can be based on several lines of argument. The most direct 
evidence is the 
observed rise of the temperature $T$ of the cosmic microwave background (CMB) 
radiation with increasing redshift $z$, i.e. $T(z) = T_0 (1+z)$ \cite{Tz}. 
Together with the observed Planck spectrum of the CMB (with 
$T_0 = 2.725\pm 0.001$ K \cite{COBE/FIRAS}\footnote{In the following we will 
set Boltzmann's constant $k=1$, and thus measure temperature in units of eV,
i.e.~$1$ K $= 8.617 \times$ $10^{-5}$~eV. Unless stated otherwise, we 
set $c=\hbar=1$.}), 
we can conclude that the radiation energy density is given by the 
Stefan--Boltzmann law, 
\begin{equation}
\epsilon = \frac{\pi^2}{30} g_\epsilon (T) T^4 \propto (1+z)^4,
\end{equation}
which, at large enough $z$, dominates the energy density of matter 
$\epsilon_{\rm m} = m n \propto (1+z)^3$. The function $g_\epsilon (T)$ 
counts the effective number of relativistic helicity degrees of freedom at 
a given photon temperature $T$ (fermionic degrees of freedom are suppressed
by a factor $7/8$ with respect to bosonic degrees of freedom). 
For times after 
$e^+ e^-$ annihilation (at $T_{e^+ e^-} \sim m_e/3 \approx 170$ keV) 
one finds 
\begin{equation}
{\epsilon \over \epsilon_{\rm m}} = 2.769 \times 10^{-4} 
\left({0.15 \over \omega_{\rm m}}\right) \left({T \over T_0} \right) = 
{1+z\over 1 + z_{\rm eq}},
\end{equation}
where $\omega_{\rm m}$ measures the mass density of 
matter\footnote{$\omega_{\rm m}$ measures the mass of non-relativistic matter 
in the Universe today, i.e.~baryons and cold dark matter. It is defined as
$\omega_{\rm m} \equiv \rho_{\rm m}/[3(H_0/h)^2/8\pi G]$, where 
$\rho_{\rm m}$ denotes the mass density and the expansion (Hubble)
rate is given by $H_0 \equiv 100 h\, \mbox{km/s/Mpc}$. Recent data from 
WMAP, combined with data from other CMB experiments and the 2dF 
galaxy redshift survey, provide $\omega_{\rm m} = 0.136\pm 0.009$ and 
$h = 0.71\pm 0.04$, within a fit to a $\Lambda$CDM model 
with running spectral index \cite{Spergel}.};
the redshift of matter--radiation equality is given by $1 + z_{\rm eq} = 
3612 (\omega_{\rm m}/0.15)$.  
Before $e^+ e^-$ annihilation we have 
\begin{eqnarray}
{\epsilon \over \epsilon_{\rm m}} &=&  1.371 \times 10^6 
\left({0.15 \over \omega_{\rm m}}\right) 
\left({g_\epsilon (T) \over 10.75}\right)\left({T\over 1 \mbox{\ MeV}}\right)
\\
&=& 0.830 \left({g_\epsilon (T) \over 10.75}\right)
    \left({1 + z\over 1 + z_{\rm eq}}\right); \\
1+z &=&  5.966 \times 10^9 \left({T\over 1 \mbox{\ MeV}}\right) \ .
\end{eqnarray}    
Thus, during the first second, radiation is dominating matter and 
all other possible components of the Universe (such as curvature, which 
scales with $a^{-2}$, or a cosmological constant, which does not change 
during the cosmic evolution). 

It is very useful to introduce the fundamental cosmological scale,
associated with a given temperature of the Universe. The Friedmann 
equation links the expansion (Hubble) rate $H$ to the mass density 
of the Universe, 
\begin{equation} 
\label{Friedmann} H^2 = {8\pi G \over 3} \rho \ , 
\end{equation} 
where $\rho \equiv \epsilon/c^2$. This equation is obtained under the 
assumption that general relativity is the appropriate description of gravity
and that the space-time is isotropic and homogeneous, with vanishing spatial 
curvature and cosmological constant. 
For dynamical questions, the Hubble time\footnote{In a radiation-dominated
Universe the cosmic time is proportional to the Hubble time, $t = t_{\rm H}/2$.
However, in this work we refer to the Hubble age, since only lower limits 
can be given for the age of the Universe; e.g.~a past infinite epoch of 
cosmological inflation (avoiding any singularity) is possible before 
the radiation-dominated epoch \cite{EM2002}.}, 
$t_{\rm H} \equiv 1/H$, 
is the typical time interval in which any of the thermodynamic variables,
the curvature and the expansion of the Universe, can change significantly:
\begin{equation}
t_{\rm H} = \left(10.75\over g_\epsilon(T) \right)^{1/2}
\left( 1 \mbox{\ MeV}\over T\right)^2 1.476 \mbox{\ s}.
\end{equation}
The time scale of the QCD transition is $10$~$\mu$s, that
of the EW transition is $10$~ps. It is straightforward to convert
these time scales into distances; the Hubble distance 
$R_{\rm H} \equiv c t_{\rm H}$ at $1 (160, 10^5)$ MeV is
$4 \times 10^5$ km ($10$ km, $10$ mm). These distances are the physical
distances at the given temperatures and they grow with the expansion.

\begin{figure}
\centerline{\includegraphics[width=0.65\textwidth]{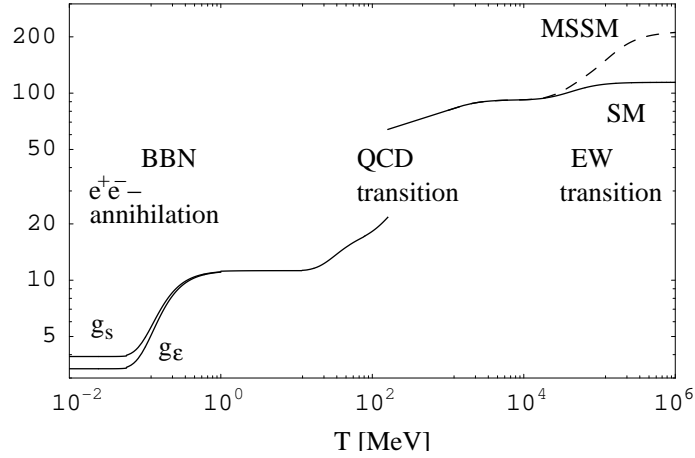}}
 \caption{The effective number of degrees of freedom 
          $g_\epsilon (T) = \epsilon(T)/(\pi^2/30 T^4)$. The full line is the 
          prediction of the standard model of particle physics, the dashed 
          line shows a minimal supersymmetric extension of the standard model
          (SPS1a) \cite{SPS}.}
 \label{fig1}
\end{figure}
Let us now consider the effective number of relativistic helicity degrees 
of freedom as a function of the temperature, see fig.~\ref{fig1}. The two 
full lines show the effective degrees of freedom of the energy density,
$g_\epsilon$ and of the entropy density $g_{\rm s}$ (upper line) for the
particle content of the standard model. In this plot the measured
particle masses are taken from the Particle Data Group \cite{PDG}. With 
increasing temperature the following important events happen: at temperatures
below $T_{e^+e^-} \sim m_e/3$, present values $g_\epsilon (T_0) = 3.363$ and
$g_{\rm s}(T_0) = 3.909$ are taken. The difference between $g_\epsilon$ and 
$g_{\rm s}$, the effective number of relativistic helicity degrees of freedom 
that contribute to the entropy density, is due to the difference of the 
photon and neutrino temperatures $T_\nu/T_\gamma = (4/11)^{1/3}$ after 
$e^+e^-$ annihilation. At temperatures above $1$ MeV, electrons,  
photons and neutrinos have the same temperature. As the 
temperature increases, particles with mass $m$ become relativistic at 
approximately $m/3$. The rise of $g$ starting at around $30$ MeV is mainly 
due to muons and pions, but also heavier hadrons can be excited. In 
figure \ref{fig1} all hadrons 
up to the $\omega$ meson ($m_\omega = 783$ MeV) have been considered 
(finite volume effects are ignored). At the temperature of the QCD transition, 
here $T_{\rm QCD} = 160$ MeV, the number of degrees of freedom changes very 
rapidly, since quarks and gluons are coloured. As discussed in more detail 
below, the order of the QCD transition is still unknown. Thus the details
of figure \ref{fig1} in the vicinity of the QCD transition are just an 
indication of what could happen. The strange mass is assumed to be $120$ MeV 
here. At still higher temperatures again heavier particles are excited, but 
within the standard model of particle physics nothing spectacular happens. 
Especially the EW transition is only a tiny effect in figure 
\ref{fig1}. (Here, we assumed a top mass of $174$ GeV and a Higgs 
mass of $120$ GeV.) This situation changes if the minimal supersymmetric 
extension of the standard model (MSSM) is considered. The effective degrees of 
freedom are more than doubled, since all supersymmetric partners 
can eventually be excited\footnote{The standard model has $g = 106.75$, 
the MSSM has $g = 228.75$, when the temperature is larger than all particle 
masses.}. For the purpose of figure~\ref{fig1}, the mass spectrum of the 
``Snowmass Points and Slopes'' model 1a \cite{SPS} has been used. Note that 
this model has light superpartners, compared with the other models. The 
lightest supersymmetric particle in the SPS1a model is the lightest neutralino
with a mass of $95$ GeV. 

Figure \ref{fig1} clearly highlights three potentially very 
interesting epochs: (i) the EW transition at $100$--$200$~GeV, 
(ii) the QCD transition at $150$--$180$~MeV, and (iii)
the $e^+e^-$ annihilation at $\sim 170$ keV. At these events the number 
of relativistic 
helicity degrees of freedom changes dramatically, and so the entropy
density of the radiation fluid. Figure \ref{fig1} is therefore very useful 
to tell us which events in the radiation fluid might be of interest from 
the thermodynamic point of view. 

$e^+e^-$ annihilation is treated in detail in many cosmology textbooks
and we refer the reader to \cite{Kolb90}. The three neutrino flavours are 
decoupled chemically and kinetically from the plasma at temperatures below 
$1$~MeV; thus the entropy of the relativistic electrons is transferred to the 
photon entropy, but not to the neutrino entropy when electrons and positrons
annihilate. This leads to an increase of the photon temperature relative 
to the neutrino temperature by $T_\nu/T_\gamma = (4/11)^{1/3}$. 

For the EW transition 
we still do not know the particle content of the Universe at the relevant 
temperature. We will not study its cosmological implications here. 
During this transition, according to the standard model of particle 
physics, all particles except the Higgs acquire their mass by the mechanism 
of spontaneous symmetry breaking. For a Higgs mass above $72$ GeV (the
current mass limit is $114.3$ GeV at $95\%$ CL \cite{PDG}), it has been 
shown that the transition is a crossover, and since the change in 
relativistic degrees of freedom is tiny, this is a very boring event from 
the thermodynamical perspective. However, extensions of the standard model
offer the possibility that the transition is of first order, which implies 
that electroweak baryogenesis might be possible in such a case (although 
the region of parameter space in which this can happen is small). 
For recent results on the order and temperature of the electroweak 
transition, see \cite{ew}.  
A cosmologically most relevant consequence of the epoch before the electroweak 
transition is that sphaleron processes allow a violation of baryon number 
$B$ and of lepton number $L$, in a way that $\Delta(B-L)=0$ \cite{KRS}. 
At high temperatures $T > T_{\rm EW}$, these processes take place at a 
rate per unit volume of the order $g^{10}\ln(1/g) T^4$ ($g$ denotes the 
SU(2) coupling) \cite{Boedeker}, which leads to a rapid equilibration 
between baryon and lepton number. This has the important consequence 
that (within the standard model) the sphaleron processes predict 
$L = - (51/28)B$. The numerical coefficient depends on the particle content 
before the electroweak transition, e.g. for two complex Higgs doublets 
$L = - (15/8)B$ \cite{Harvey90}. In any case the lepton asymmetry has to be 
of the same order of magnitude as the baryon asymmetry. This is 
important for the question whether chemical potentials can play a role in 
the early Universe. We show below (section \ref{baryons}) that they are 
completely irrelevant for the present purpose. 

Thus, from the thermodynamic point of view, the QCD epoch is the most 
interesting event during the first second. However, not all interesting 
physics is covered by thermodynamics. Especially phenomena that affect
the baryon content and cold dark matter are not only governed by the
bulk thermodynamics. We therefore take a closer look at the interaction
rates. 

\begin{figure}
\centerline{\includegraphics[width=0.65\textwidth]{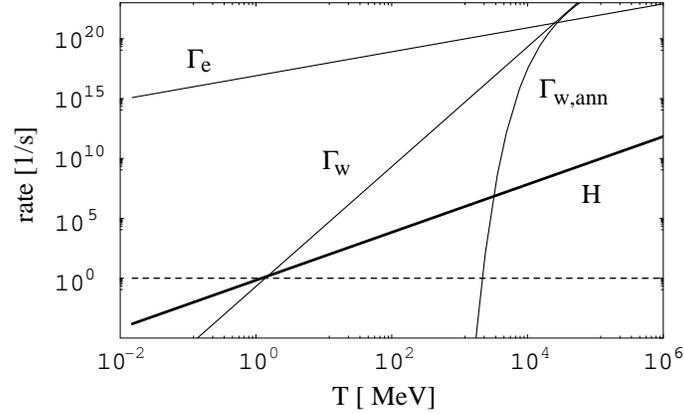}}
\caption{The Hubble rate $H$ and the typical interaction rates of weak 
          ($\Gamma_{\rm w}$) and electric ($\Gamma_{\rm e}$) processes that 
          involve relativistic particles, as well as the typical 
          rate of a weak annihilation rate $\Gamma_{\rm w,ann}$ for a 
          particle mass of $100$ GeV. At $t_{\rm H} \sim 1$ s, the weak 
          interaction
          rate falls below the expansion rate (chemical and kinetic decoupling 
          of neutrinos, kinetic decoupling of neutralinos), at temperatures
          of the order of $1$--$10$ GeV neutralinos freeze-out. The electric 
          interaction rate stays well above the Hubble rate up to the 
          epoch of photon decoupling, well after the epoch we are interested in
          here. The dashed line indicates the rate $1/$s. (Note that 
          for the purpose of this sketch the weak interactions are 
          approximated by the Fermi theory, which breaks down 
          at $T \sim m_W/3$. We also ignored the running of the coupling 
          constants in this figure.)}
 \label{fig2}
\end{figure}

A complementary picture of the early Universe arises from a comparison 
of the most relevant interaction rates and the expansion rate, sketched 
in figure \ref{fig2} as a function of temperature. Strong, electric, and weak 
interactions keep all relativistic particles in kinetic and chemical 
equilibrium down to temperatures of $\sim 1$ MeV. At that point neutrinos and
neutrons decouple chemically and kinetically from the rest of the radiation 
fluid. This has several important implications: (i) During the process of 
neutrino decoupling, density inhomogeneities of the radiation fluid
on scales below $1/30$ of the Hubble distance at $1$ MeV are washed out by 
collisional damping. Irrespectively of the initial conditions of the Universe,
this guarantees that entropy is distributed homogeneously within large patches,
and thus justifies neglecting temperature fluctuations on small scales during 
big bang nucleosynthesis (BBN).    
(ii) The weak equilibrium between neutrons and protons freezes out as the 
neutrinos decouple. The neutron-to-proton ratio changes only because of neutron 
decay after weak decoupling. 
(iii) Neutrons no longer scatter and can travel ``large'' distances. They 
can smear out local fluctuations of the neutron-to-baryon ratio. 

Because of neutrino oscillations, the individual neutrino flavour is not 
conserved, but a flavour equilibrium is established at a time scale of the 
order of the oscillation time, as soon as the radiation fluid is dilute 
enough to allow a sufficient mean free path for the neutrinos. For a scenario 
based on the atmospheric and solar neutrino data, favouring large mixing 
angles, it was shown by Dolgov et al. \cite{Dolgov02} that flavour equilibrium 
is established at about $T \sim  3$ MeV. 

Weakly interacting massive particles (WIMPs) are excellent candidates for the
CDM. A prominent example is the lightest supersymmetric particle (LSP), most 
likely the lightest neutralino. WIMPs are exponentially suppressed as the 
temperature drops below their mass. However, since they interact only weakly,
the annihilation rate cannot keep up with the Hubble rate and they drop 
out of chemical equilibrium (freeze-out). This happens at typically 
$T \sim m/20$. In figure \ref{fig2}, $m = 100$ GeV. Nevertheless, they are kept
in kinetic equilibrium down to temperatures of $10$--$100$ MeV \cite{HSS} by 
elastic scattering.
During and after their kinetic decoupling, collisional damping and free 
streaming wash out density inhomogeneities in CDM on very small scales. 
This effect might be of relevance for structure formation and the search for 
dark matter.

The notation used in this review is summarized in two tables at the end
of the paper.

\section{The cosmic QCD transition: an overview \label{QCDoverview}}

One of the most spectacular epochs in the early Universe is the QCD 
epoch. QCD describes the strong interactions 
between quarks and gluons and is well tested in the perturbative regime, 
i.e. at high energies and momenta. At low energies, quarks and gluons are 
confined into hadrons. The scale of QCD is $\Lambda_{\rm QCD} = 
{\cal O}(100 \mbox{\ MeV})$. At temperatures $T \sim \Lambda_{\rm QCD}$,
there is a transition between a quark--gluon plasma (QGP) and a hadron gas 
(HG) \cite{QCDtrans}. 
These theoretical expectations are in agreement with findings from the
CERN heavy ion programme \cite{newstate} and the ongoing studies at the 
Relativistic Heavy Ion Collider (RHIC) \cite{RHICrev} at BNL. 
 
The first studies of the cosmological QCD 
transition date back to the early 80's \cite{earlywork}. It was then realized
that a first-order QCD transition proceeds via bubble nucleation 
\cite{Hogan83,DeGrand} with small supercooling \cite{Witten,Applegate}.
The bubbles were shown to grow most probably by weak deflagration
\cite{Gyulassy,Kurki-Suonio}. Most of the following work has been based on this
scenario of a first-order QCD transition with homogeneous bubble nucleation
and bubble growth by weak deflagration. A detailed discussion of this 
scenario has been given by Kajantie and Kurki-Suonio \cite{Kajantie}.

Based on this scenario a separation of cosmic phases has been suggested 
by Witten \cite{Witten}, which may give rise to large inhomogeneities in the 
distribution of baryons in the Universe. Originally it was hoped that the 
mean bubble nucleation distance is a reasonable fraction of the size of the 
Universe, the Hubble scale. From recent lattice QCD results for the latent 
heat and surface tension in the quenched approximation (no dynamical quark 
flavours) \cite{Iwasaki2}, a tiny mean bubble nucleation distance follows 
\cite{Fuller,Ignatius,Christiansen}. 

The mean bubble nucleation distance might be larger for other bubble 
nucleation scenarios. Nucleation at impurities has been investigated
by Christiansen and Madsen \cite{Christiansen}. Another possibility is that
the nucleation of bubbles is determined by pre-existing inhomogeneities of 
temperature (e.g. generated during inflation) \cite{IS}, or by pre-existing
baryon number inhomogeneities \cite{Sanyal}.

Apart from the question `What happened at the cosmological QCD transition?',
the cosmological QCD transition is of interest for two more reasons:
firstly, the QCD epoch sets the initial conditions for one of the pillars 
of standard cosmology --- big bang nucleosynthesis. The QCD 
transition and baryon transport thereafter determine the distribution of 
protons and neutrons at the beginning of nucleosynthesis. It was
first recognized by Applegate and Hogan \cite{Applegate} that an inhomogeneous
distribution of baryons due to a first-order QCD transition could
change the primordial abundances of the light elements. This issue has 
attracted a lot of interest \cite{IBBN,Mathews,Kainulainen,JR} 
because inhomogeneous BBN might allow a higher baryon-to-photon ratio than 
homogeneous BBN \cite{BBN}. However, today this possibility is ruled out 
by observations of CMB temperature anisotropies \cite{cmb,WMAP,Spergel}, which 
confirm independently the value of $\omega_{\rm b}$ (defined  
as $\omega_{\rm m}$, but for the baryon mass density)  
that is extracted from the 
theory of homogeneous BBN and the observed light element abundances\footnote{
$\omega_{\rm b} = 0.022\pm 0.001$ from a seven-parameter fit to
CMB and large scale structure data \cite{Spergel}, 
whereas $\omega_{\rm b} = 0.0205\pm 0.0018$ from the most recent 
determination of the primordial deuterium abundance \cite{bbn}.}. However, with 
the increasing precision of the determination of abundances, 
we might find ourselves in a situation in which some tension arises between the 
values of $\omega_{\rm b}$ as extracted from the individual measurements of
$^4$He, D, and $^7$Li; see \cite{Li,Cyburt} for an indication that there 
might be a conflict between the observed D abundance, which is consistent
with the CMB measurements, and $^4$He and $^7$Li measurements. Besides 
systematic errors of the abundance determination, this might be 
a signature for an inhomogeneous BBN. 
Moreover, an inhomogeneous BBN might allow the primordial production of 
heavy elements such as $^{12}$C \cite{JedamzikHE}. 

Secondly, the QCD transition might generate relics that might be observable 
today. Most of these relics are only formed if the QCD transition is 
first-order and if the latent heat is much larger than that found
from quenched QCD in lattice calculations. Strange quark nuggets as dark 
matter and gravitational waves from colliding bubbles have been suggested 
by Witten \cite{Witten}. Hogan \cite{Hogan83b} and more recently Cheng and 
Olinto \cite{Olinto} argued that magnetic fields might be generated. 
Today, these relics appear to be unlikely, because 
the typical scale for bubble nucleation is very small. 

Recently, Zhitnitsky \cite{Zhitnitsky02} suggested a new CDM candidate:
QCD balls. If axions existed and if the reheating scale after inflation 
lied above the Peccei--Quinn scale, collapsing axion domain walls could 
trap a large number of quarks. At some point the collapse would be stopped 
by the Fermi pressure of the quarks, which would settle in a colour 
superconducting phase \cite{Alford98}. This process takes place during the QCD 
transition, but does not require a first-order transition, contrary to the 
idea of strange quark nuggets. Brandenberger, Halperin, and 
Zhitnitsky \cite{Brandenberger} speculated that even a separation of 
baryons and antibaryons due to the non-perturbative QCD vacuum might be 
possible during the QCD epoch. This would effectively give a new baryogenesis 
scenario and a new candidate for dark matter. It seems to me that much more 
work has to be done to explore these exciting ideas.  
According to the mentioned suggestions, in the best of all QCD worlds, it 
might be possible to explain baryogenesis and the nature of CDM!  

It was speculated for various reasons that black holes could form 
during the QCD transition \cite{Crawford,Hall,Jedamzik}. From the present-day 
perspective these scenarios seem to be highly unlikely.

The QCD transition itself might also lead to the formation of small CDM 
clumps \cite{SSWHA,SSW,SSW2}. The speed of sound
vanishes during a first-order QCD transition and thus the restoring forces 
vanish. This leads to large amplifications of primordial density fluctuations 
in the radiation fluid and in cold dark matter. This mechanism can work for
axions or primordial black holes, since they are kinetically decoupled 
at the QCD transition, but not for WIMPs, as they belong to the radiation
fluid during the transition.

Independently from the order of the transition, a primordial background of 
gravitational waves is modified by the QCD transition, as shown in 
\cite{Schwarz}. 

Reviews on the cosmological QCD transition may be found in
Refs.~\cite{ref1,ref2,ref3,ref4,ref5}.

\subsection{Scales}

The QCD transition is expected to take place at $T_\star \sim 
\Lambda_{\rm QCD}$. We would like to measure $T_\star$ from heavy ion 
collisions, but it turns out that this is not a simple task. A temperature 
estimate can be obtained from the study of measured hadron abundances 
in reletivistic heavy ion collisions. After some short initial phase one 
expects that a quark--gluon plasma (QGP) is formed, which expands and 
eventually makes a thermal confinement transition. At some point the 
hadron gas is dilute enough such that the abundance of hadron species 
is fixed. At RHIC experiments, this tempertature of the chemical 
decoupling of hadrons has been estimated to be $T = 174\pm 7$~MeV at a   
baryon chemical potential $\mu_{\rm B} = 46\pm 5$~MeV (statistical errors 
only) \cite{BraunMunzinger}. This reasoning suggest that the QCD transition 
temperature should lie above the estimated hadron freeze-out temperature 
and thus $T_\star > 160$~MeV ($95$\% C.L.). Although the   
baryon chemical potential at RHIC is small with respect to previous heavy ion 
experiments, it is huge from the point of view of cosmology. However,
lattice QCD predicts a suprisingly small curvature of the transition 
temperature as a function of the baryon chemical potential at $\mu_{\rm B}=0$
(for $2+1$ quark flavours, see \cite{Fodor}). Thus, the lower limit on the
transition temperature suggested by RHIC should be applicable in cosmology.
Another difference between heavy ion collisons and cosmology should be stressed:
while the time scale of the cosmological QCD transition is $10^{-5}$ s,
it is $10^{-23}$ s in the laboratory. It is therefore necessary to check 
the theoretical and experimental estimates of $T_\star$ by lattice 
QCD `experiments'.

{}From recent lattice QCD calculations for quenched QCD (no dynamical quarks)
the transition temperature is $T_\star \approx 271\pm 2$ MeV 
\cite{Boyd,CP-PACSquenched,KarschRev}. For two-flavour QCD 
$T_\star \approx 171\pm 4$~MeV \cite{KLP,CP-PACS,CP-PACS2flavour,KarschRev}, 
whereas for three-flavour QCD $T_\star \approx 154\pm 8$ MeV 
\cite{KLP,KarschRev}, almost independent of the quark mass. For the most 
interesting case of two light quark flavours (up and down) and one massive 
strange quark, no values for the transition temperature have been obtained 
so far. In the following I pick a transition temperature of 
$T_\star = 160$~MeV, i.e. I implicitly assume that the physical situation 
resembles more closely the three-flavour case than the two-flavour situation.
Recent reviews of thermal QCD simulations can be found in \cite{KarschRev}.

\subsubsection{The Hubble scale \label{Hubble}}

It is a good approximation for our purpose to treat all particles with 
$m \ll 3 T$ as though they were massless; all other particles 
are neglected in the total energy density $\epsilon$.
Above the QCD transition $g_{\rm quarks} = (7/8) 12 N_f$ (number of 
quark flavours) and 
$g_{\rm gluons} = 8$, below the QCD transition $g_{\rm pions} = 3$. 
At the QCD epoch there are the photons, three flavours of neutrinos, and
electrons and muons. Counting all up we find for the QCD epoch
\begin{equation}
g(T > T_\star) = 51.25\, (61.75) \ ,
\end{equation}
without (with) strange quarks, and 
\begin{equation}
g(T < T_\star) = 17.25\, (21.25)\ ,
\end{equation}
without (with) kaons.

At the QCD transition the Hubble radius is about $10$ km: 
\begin{equation}
R_{\rm H} \approx \left( 61.75\over g \right)^\frac12 
                  \left(160 \mbox{\ MeV} \over T_\star\right)^2
                  7.2 \mbox{\ km} \ ,
\end{equation}
before the transition, and 
\begin{equation}
R_{\rm H} \approx \left( 21.25\over g \right)^\frac12
                  \left(160 \mbox{\ MeV} \over T_\star\right)^2
                  12 \mbox{\ km} 
\end{equation}
after it.

Today this corresponds to scales of $1$ pc or $3$ light-years.
The Hubble time at the QCD transition, $t_{\rm H} \sim 10^{-5}$ s, is 
extremely long in comparison with the relaxation 
time scale of the strong interactions, which is about $1 \mbox{\ fm}/c \sim 
10^{-23}$ s. Thus, the transition is very close to an equilibrium process. 

The mass inside a Hubble volume is $\sim 1 M_\odot$:
\begin{equation}
M_{\rm H} \equiv {4\pi\over 3} R_{\rm H}^3 \rho \approx 
                  \left( 61.75\over g \right)^\frac12
                  \left(160 \mbox{\ MeV} \over T_\star\right)^2
                  2.5 M_\odot \ .
\end{equation}
This mass is redshifted $\propto (1+z)$ as the Universe expands, 
because it is made up of radiation. 
An invariant mass is the mass of cold dark matter in a comoving volume, 
$M_{\rm cdm} \equiv (1+z_{\rm eq})/(1+z) M(z)$. At the QCD transition 
$(1+z_{\rm eq})/(1+z) \sim 10^{-8}$, and thus 
\begin{equation}
M_{\rm H}^{\rm cdm} \sim 10^{-8} M_\odot \ .
\end{equation}

Another figure of interest is the baryon number inside the Hubble volume
at the QCD transition, $B_{\rm H} \equiv (4\pi/3) R_{\rm H}^3 n_{\rm B}$. 
The baryon number density $n_{\rm B}$ follows from the ratio
of baryons to photons at BBN, $\eta \equiv (n_{\rm B}/n_\gamma)_{\rm BBN}$,
and the conservation of baryon number and entropy, $n_{\rm B}/s = $ constant: 
\begin{equation}
n_{\rm B}(T_\star) = \eta\left(n_{\gamma}\over s\right)_{\rm BBN} s(T_\star) \ ,
\end{equation}
Finally, the baryon number inside a Hubble volume reads:
\begin{equation}
B_{\rm H} \approx  \left(61.75\over g \right)^\frac12
             \left(160 \mbox{\ MeV} \over T_\star \right)^3 
             \left(\eta\over 5.6\times  10^{-10}\right) 2.1 \times 10^{48}  
\end{equation}
at the beginning of the transition and about twice that value at the 
end\footnote{This formula is correct if no black holes 
are formed during the QCD transition and if the quark nuggets
that might have formed evaporate before the BBN epoch.}.
{}From standard BBN and the most recent deuterium measurements,
$\eta = (5.6\pm 0.5) \times 10^{-10}$ \cite{bbn}.
\begin{figure}[t]
\centerline{\includegraphics[width=0.4\textwidth]{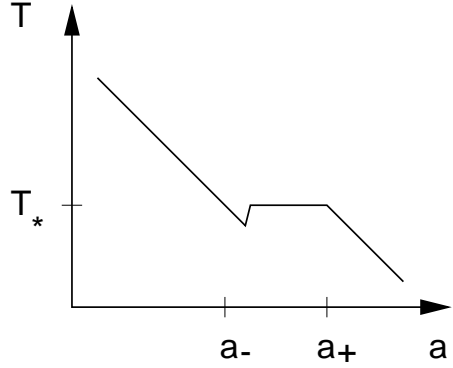}}
\caption{Qualitative behaviour of the temperature $T$ as a function of the 
scale factor $a$ during a first-order QCD transition with small supercooling. 
Above the critical temperature $T_\star$ the Universe (filled with a 
quark--gluon plasma) cools down thanks to its expansion ($a<a_-$). After a 
tiny period 
of supercooling (in the figure the amount of supercooling and its duration 
are exaggerated) bubbles of the new phase (hadron gas) nucleate. Almost in 
an instant these bubbles release enough energy (latent heat) to reheat the 
Universe to the critical temperature. During the rest of the transition 
both phases coexist in pressure and temperature equilibrium ($a_- < a < a_+$). 
Therefore the temperature is constant. After reheating the bubble growth is 
governed by the expansion of the Universe. The transition is completed when 
no quark--gluon plasma is left over. For $a>a_+$ the temperature 
decreases again due to the expansion of the Universe.
\label{fig3}}
\end{figure}

\subsubsection{The bubble scale}

The typical duration of a first-order QCD transition is $0.1 t_{\rm H}$. 
If the cosmological QCD transition is of first order, it proceeds via 
bubble nucleation. 
{}From the small values of surface tension and latent heat found in 
lattice QCD calculations \cite{Iwasaki2}, the amount of supercooling 
is found to be small \cite{Ignatius}. The temperature as a function of the 
scale factor is shown in fig.~\ref{fig3} for the small supercooling scenario.
Hadronic bubbles nucleate during a short period of supercooling, 
$\Delta t_{\rm sc} \sim 10^{-3} t_{\rm H}$. The typical bubble nucleation 
distance is
\begin{equation}
d_{\rm nuc} \sim 1 \mbox{\ cm\ } \sim 10^{-6} R_{\rm H}
\end{equation}
for homogeneous nucleation \cite{Christiansen}. 
The hadronic bubbles grow very fast, within $10^{-6} t_{\rm H}$, until 
the released latent heat has reheated the Universe to $T_\star$. By that 
time, just a little fraction of volume has gone through the transition.
For the remaining $99\%$ of the transition, the HG and the QGP
coexist at the pressure $p_{\rm HG}(T_\star) = p_{\rm QGP}(T_\star)$. During
this time the hadronic bubbles grow slowly and the released latent heat keeps
the temperature constant until the transition is completed. The energy density
decreases continuously from $\epsilon_{\rm QGP}(T_\star)$ at the beginning 
of the transition to $\epsilon_{\rm HG}(T_\star)$ when the transition is 
completed.

\subsection{Order of the thermal QCD transition}

The order of the QCD phase transition and (for a first-order transition) the
magnitude of the latent heat are still a subject of debate. Below I first
introduce the bag model, because it gives a simple parametrization for the 
pressure $p(T)$ and it is useful to introduce some notation. Then I sum
up the knowledge obtained from lattice QCD (order of the transition, 
latent heat, surface tension). 

\subsubsection{The bag model \label{bagm}}

The MIT bag model \cite{bag}
represents the short-distance dynamics by an collisionless gas of massless 
quarks and gluons [in the following I call that a Stefan--Boltzmann (SB) gas 
for brevity] and the long-distance confinement effects
by a constant negative contribution to the pressure, the bag constant $B$,
\begin{equation}
\label{pqgp}
p_{\rm QGP}(T) = p_{\rm QGP}^{\rm SB}(T)  - B \ ,
\end{equation}
with $p_{\rm QGP}^{\rm SB}(T)=(\pi^2/90) g_{\rm QGP} T^4$. Here I take 
$g_{\rm QGP} = 47.5$.
The low-temperature phase is a hadron gas. It may be modelled as an
SB gas of massless pions and kaons (since $m_K \sim 3 T_\star$), 
$g_{\rm HG} = 7$,  
\begin{equation}
\label{phg}
p_{\rm HG} = {\pi^2\over 90} g_{\rm HG} T^4.
\end{equation}

At the phase transition the pressures of the quark--gluon phase and the hadron
phase are in equilibrium, 
\begin{equation}
p_{\rm QGP}(T_\star) = p_{\rm HG}(T_\star) \equiv p_\star \ . 
\end{equation}
This condition gives, together with (\ref{pqgp}) and (\ref{phg}), the relation 
between $T_{\star}$ and $B$:
\begin{equation}
\label{Tstar}
B = {\pi^2\over 90} (g_{\rm QGP} - g_{\rm HG}) T_\star^4 \ .
\end{equation}
Let me take $T_\star$ from lattice QCD calculations with two and three 
flavours,  which indicate $T_\star =  150$ to $180$~MeV \cite{KarschRev}. 
This corresponds to a range of bag constants $B^{1/4} = 218$ to
$261$~MeV. This range overlaps well with fits to the light-hadron masses, 
which yield $B^{1/4} = 145$ to $245$~MeV (a compilation of various bag model
light-hadron fits can be found in \cite{Farhi}). 

The QCD transition is first order in the bag model, because the entropy density
$s \equiv {\rm d} p /{\rm d} T$ makes a jump. The latent heat per unit volume 
\begin{equation}
l \equiv T_{\star} \Delta s
\end{equation}
measures the amount of `internal' energy that is released during the phase 
transition. In the bag model the latent heat is
\begin{equation}
l = {2 \pi^2 \over 45} \Delta g T_{\star}^4 = 4 B,
\end{equation}
where $\Delta g \equiv g_{\rm QGP} - g_{\rm HG}$.

Besides the latent heat, the surface tension is the crucial parameter for the 
nucleation of bubbles (see section \ref{bubbles}). The surface tension 
\begin{equation}
\sigma \equiv \left({\rm d} W\over{\rm d} A\right)_V
\end{equation}
is the work ${\rm d} W$ that has to be done per area ${\rm d} A$ to change 
the phase interface at fixed volume. The absence of surface excitations 
in hadronic spectra suggests that $\sigma^{1/3} \ll B^{1/4}$ \cite{Farhi} 
in the bag model. This implies $\sigma \ll 3.1 T_\star^3$ for three quark 
flavours. A self-consistent calculation of the surface tension within the 
MIT bag model shows that the surface tension vanishes for massless quarks 
and gluons if no interactions besides the bag constant (i.e. $\alpha_s = 0$) 
are taken into account \cite{Farhi}. This can be cured by introducing
short-range interactions \cite{Farhi} or by including the strange quark with 
mass \cite{surface-tension}, $m_s = 80$ to $150$ MeV \cite{PDG}. 

\subsubsection{Lattice QCD results}

In quenched QCD the phase transition is of first order \cite{Boyd}. The 
latent heat was determined to be $l \approx 1.4 T_\star^4$ \cite{Iwasaki2}. 
It is useful to take the ratio $R_L$ of the latent heat to the value 
$T_{\star} \Delta s^{\rm SB}$, where $\Delta s^{\rm SB}$ is 
the difference in entropy between two SB gases:
\begin{equation}
R_L \equiv {l\over (T_\star \Delta s)^{\rm SB}} =
\left\{ \begin{array}{ll}
               1   & \mbox{bag model} \\
               0.2 & \mbox{quenched lattice QCD}
        \end{array}\right. \ .
\end{equation}
For two light quarks it is likely that the transition is a crossover
\cite{MILC96,MILC97}. This is in agreement with theoretical considerations
\cite{Wilczek}, which predict a second-order phase transition in the massless
quark limit. 

For three flavours close to the chiral limit, the phase transition is again 
of first order and some simulations suggest that this holds true for the 
physical case \cite{Iwasaki}. The latter result was obtained using the 
Wilson quark action, whereas results with staggered quarks \cite{Brown,KLP} 
indicate a crossover for the physical quark masses. For four quark flavours 
the transition is first order \cite{Karsch97}. 
For a detailed reviews on these issues, see \cite{KarschRev}.

Since the latent heat for lattice QCD is known for quenched QCD only, 
I decided to use the latent heat ratio $R_L = 0.2$ from quenched QCD as an 
indication for the physical case.

For the surface tension rather small values are found from lattice QCD. 
It is reported in \cite{Iwasaki2} that $\sigma \approx 0.015 T_\star^3$ for 
quenched lattice QCD. There are no values for unquenched QCD available yet. 
However, an upper bound was obtained for the case of four-flavour lattice QCD 
in Ref.~\cite{Markum}, i.e. $\sigma < 0.1 T^3_\star$. 

\subsection{Effects from a first-order QCD transition}

Let me now briefly summarize the effects that have been suggested to
emerge from the cosmological QCD transition. There are two kinds of effects:
the effects that have been found in the mid 80s and early 90s 
stem from the bubble scale and they thus affect scales 
$\lambda \leq d_{\rm nuc}$. The formation of quark nuggets, the generation 
of isothermal baryon fluctuations, the generation of magnetic fields and 
gravitational waves belong to the effects from the bubble scale. 

In recent years it was found that there is another class of possible 
consequences from the QCD transition, which are connected to the Hubble 
scale and therefore affect scales $\lambda \leq R_{\rm H}$. Among these effects
are the amplification of inhomogeneities and later formation of cold dark matter
clumps, the modification of primordial gravitational waves, and the enhanced 
probability of black hole formation during the QCD transition. 

\subsubsection{Quark nuggets/Strangelets}

In the mid 80s interest in the cosmological QCD transition arose, because
it was realized that a strong first-order QCD phase transition could lead to
important observable signatures. Most of the interest was based on a 
separation of
cosmic phase as suggested by Witten \cite{Witten}. If the cosmological 
QCD transition is first-order, bubbles of hadron gas are nucleated and grow
until they merge and fill up the whole Universe. Towards the end of 
the coexistence of the QGP and HG phases, shrinking quark droplets remain.
These droplets are expected to be baryon-enriched with respect to the hadron
phase since in equilibrium, baryons are suppressed by the Boltzmann factor
in the HG, and baryon diffusion across the phase boundary might be inefficient. 

In 1971 Bodmer \cite{Bodmer} suggested the possibility that strange quark 
matter might be the ground state of bulk matter, instead of ${}^{56}$Fe. 
Later Witten \cite{Witten} discovered this idea again. Strange quark matter 
was further studied by Farhi and Jaffe \cite{Farhi}.  
The idea of strange quark matter is based on the observation that the 
Pauli principle allows more quarks to be packed into a fixed volume in phase
space if three instead of two flavours are available. Thus the energy per 
baryon would be lower in strange quark matter than in nuclei. 
However, the strange quark is heavy compared with up and down quarks, and 
this mass counteracts the advantage from the
Pauli principle. No strange quark matter has been found experimentally so far
\cite{smexp}. The issue of stability of strange quark matter has 
not been settled yet; for a recent review see \cite{Madsen}.

Witten \cite{Witten} pointed out that a separation of phases during the 
coexistence of the hadronic and the quark phase could gather a large number 
of baryons in strange quark nuggets \cite{Witten}. 
These quark nuggets could contribute to the dark matter today \cite{Witten}
or affect BBN \cite{smBBN}. At the end of the transition the baryon number 
in the quark droplets could exceed the baryon number in the hadron phase 
by several orders of magnitude, $n_{\rm B}^{\rm QGP}$ could be close to 
nuclear density \cite{Sumiyoshi}. However, it was realized
that the quark nuggets, while cooling, lose baryons. The quark nuggets
evaporate as long as the temperature is above $\sim 50$~MeV \cite{Alcock}. 
Quark nuggets may survive this evaporation if they contain much more 
than $\sim 10^{44}$ baryons initially \cite{evap}. This number should be 
compared with the number of baryons inside a Hubble volume at the QCD 
transition, which is $10^{48}$ (see section \ref{Hubble}). Thus, the mean 
bubble nucleation distance should be $> 3 \times 10^{-2} R_{\rm H} 
\sim 300$ m so as to collect enough baryons. This seems impossible from 
recent lattice QCD calculations of latent heat and surface tension 
\cite{Iwasaki2}. 

In \cite{Sumiyoshi,evap} a chromoelectric flux-tube model was used to 
estimate the penetration rate of baryons through the interface.
A quark that tries to penetrate the interface creates a flux tube, which
most probably breaks up into a quark--antiquark pair. By this mechanism,
mesons evaporate easily. On the other hand, baryons are rarely formed,
because a diquark--antidiquark pair has to be produced in the break up
of the flux tube. However, one could think of mechanisms that would increase 
the evaporation rate of baryons.
If a significant fraction of diquarks was formed in the quark phase,
these diquarks could penetrate the interface by creating a flux tube, which
eventually breaks, creating a quark--antiquark pair. Then the quark would 
evaporate together with the diquark and form a baryon, whereas the antidiquark
would remain in the quark phase. Such a mechanism would increase the 
evaporation rate. 
Thus, independently from the existence (stability) of strange quark matter 
it seems highly unlikely that strange quark nuggets could survive after 
the cosmological QCD transition below temperatures of $\sim 50$ MeV.

However, see Bhattacharyya et al. \cite{Bhattacharyya00} for a different 
point of view. They conclude that stable quark nuggets could be formed if 
the QCD transition is a strong first-order transition (they use the bag model)
and if the critical temperature is around $100$ MeV, instead of $150$--$180$ 
MeV as indicated  by lattice QCD. They speculate that these quark nuggets 
could account for all the CDM. In a recent work \cite{Banerjee02}, it was 
suggested that these primordial quark nuggets might clump by gravitational 
attraction and eventually form bound objects of $0.5 M_\odot$, which would 
explain the gravitational microlensing events that have been observed 
towards the Large Magellanic Cloud \cite{microlensing}.

\subsubsection{Inhomogeneous nucleosynthesis}

Applegate and Hogan \cite{Applegate} found that a strong first-order 
QCD phase transition induces inhomogeneous nucleosynthesis. It is extremely
important to understand the initial conditions for BBN, because many of
our ideas about the early Universe rely on the validity of the standard
(homogeneous) BBN scenario. This is in good agreement 
with observations \cite{BBN}. In inhomogeneous nucleosynthesis \cite{IBBN},
large isothermal fluctuations of the baryon number (the remnants of the
quark droplets at the end of the QCD transition) could lead to different
yields of light elements. As a minimal requirement for an inhomogeneous 
scenario of nucleosynthesis, the mean bubble nucleation distance
has to be larger than the proton diffusion length, which corresponds
to $\sim 3$ m \cite{Mathews} at the QCD transition. This is two orders
of magnitude above recent estimates of the typical nucleation distance
\cite{Christiansen}.

On the other hand the observed cosmic abundances of light elements
do not favour inhomogeneous nucleosynthesis, except a small region in
parameter space corresponding to an inhomogeneity scale of $\sim
40$ m \cite{Mathews}.

However, interesting inhomogeneity scales for BBN might follow if the bubble
nucleation is not homogeneous. Heterogeneous nucleation in the presence 
of impurities or nucleation in the inhomogeneous Universe could increase
the baryon inhomogeneity scale to interesting values.

Although values for $\eta$ dramatically different from those in 
the standard BBN are excluded both from measurements of the light element 
abundances and from the CMB, it might be possible to alter 
the primordial abundance of heavy elements ($A > 7$) very much in various 
inhomogeneous scenarios \cite{JedamzikHE}. More on inhomogeneous BBN will 
be discussed in section \ref{baryonfluc}.

\subsubsection{Cold dark matter clumps}

Scales $\lambda$ that are of the order of the Hubble radius $R_{\rm H}$ are
not sensitive to details of the bubbles.
It was reported in Refs.\ \cite{SSWHA,SSW,SSW2} that the evolution of 
cosmological density 
perturbations is strongly affected by a first-order QCD transition for 
subhorizon scales, $\lambda < R_{\rm H}$. Cosmological perturbations on 
all scales are predicted by inflation 
\cite{perturbations,BST,Bardeen,Mukhanov,Martin},   
observed in the temperature fluctuations of the cosmic microwave 
background radiation, first by the COBE satellite \cite{Smoot,COBE}.

In the radiation-dominated Universe subhorizon density perturbations
perform acoustic oscillations. The restoring force is provided by
pressure gradients. These, and therefore the isentropic 
speed of sound $c_s = \left(\partial p/\partial \epsilon \right)_S^{1/2}$
(on scales much larger than the bubble separation scale) drop to
zero at a first-order QCD transition \cite{SSW}, because both phases
coexist at the pressure $p_\star$ only ($a$ is the scale factor of
the Universe):
\begin{equation}
c_s^2 = {{\rm d} p_\star/{\rm d} a\over {\rm d} \epsilon(a)/{\rm d} a} = 0 \ .
\end{equation}
It stays zero during the entire transition and suddenly rises back to 
the radiation value $c_s=1/\sqrt{3}$ after the transition. A significant 
decrease in the effective speed of sound $c_s$ during the cosmological QCD 
transition was also pointed out by Jedamzik \cite{Jedamzik}.

As the speed of sound drops to zero, the restoring force for acoustic 
oscillations vanishes and density perturbations for subhorizon modes 
fall freely. The fluid velocity stays constant during this free fall. 
Perturbations of shorter wavelengths have higher velocities
at the beginning of the transition, and thus grow proportional to the wave
number $k$ during the phase transition. The primordial Harrison--Zel'dovich
spectrum \cite{HZ} of density perturbations is amplified on subhorizon scales.
The spectrum of density perturbations on superhorizon scales, 
$\lambda > R_{\rm H}$, is unaffected. At $T\sim 1$ MeV the neutrinos 
decouple from the radiation fluid. During this decoupling the large peaks 
in the radiation spectrum are wiped out by collisional damping \cite{Weinberg}.

Today a major component of the Universe is dark matter, most likely CDM. 
If CDM is kinetically decoupled from the radiation fluid
at the QCD transition, the density perturbations in CDM do not suffer
from the neutrino damping. This is the case for primordial black
holes or axions, but not for supersymmetric dark matter.
At the time of the QCD transition the energy density of CDM is small, i.e.\
$\epsilon_{\rm cdm}(T_\star) \sim 10^{-8} \epsilon_{\rm rad}(T_\star)$.
CDM falls into the potential wells provided by the dominant
radiation fluid. Thus, the CDM spectrum is amplified on subhorizon
scales. The peaks in the CDM spectrum go non-linear
shortly after radiation--matter equality. This leads to the
formation of CDM clumps with
mass $< 10^{-10} M_\odot$. Especially the clumping of axions has important
implications for axion searches \cite{Sikivie}.
If the QCD transition is strong enough, these
clumps could be detected by gravitational femtolensing \cite{femtolensing}.

\subsubsection{Other effects}
\smallskip

Generation of magnetic fields 

Hogan \cite{Hogan83b} argued that magnetic fields might be generated in violent
processes at the bubble scale. Later on field lines should reconnect  
to yield large scale magnetic fields which could play an important role 
in structure formation after the recombination of matter.  
Cheng and Olinto \cite{Olinto} suggested that such magnetic fields might 
be generated by currents on the bubble surfaces.

If the shrinking QGP droplets are baryon-enriched, there is a positive  
net charge on the inner side of the surface of the bubble walls, 
which is compensated by a negative net charge of electrons outside the 
bubble. This is due to the different Debye screening lengths of electrons 
and baryons. Thus surface currents on the bubble surfaces are possible, 
which could give rise to magnetic fields. The typical scale at their creation 
would be the bubble scale. 

Subhorizon ($\lambda < R_{\rm H}$) fields are damped during neutrino 
and photon decoupling \cite{magdamp}. In the most optimistic scenario, 
magnetic fields of the order of $10^{-20}$ gau\ss\ at the $10$ Mpc scale 
would be possible today \cite{Sigl}. 

It might be that the QCD transition leads to the generation of 
pion-strings,  predicted by an effective description of 
hadronic matter within the linear sigma model. These defects are unstable 
and decay eventually, but could seed magnetic fields  
\cite{Zhang}. The source of the magnetic field would come from the 
Adler-Bell-Jackiw anomaly, which couples the $\pi^0$ to photons. 

A quite different mechanism has been proposed in \cite{Forbes}. It relies 
on several untested assumptions, especially on the existance of 
axion domain walls and an inverse cascade 
mechanism to generate magnetic fields on scales much larger than $1$ pc today. 
\smallskip

\noindent Generation of gravitational waves 
\nopagebreak

A measure for the energy density in gravitational waves is the fractional
energy density per logarithmic frequency interval $\Omega_{\rm gw}(f)$
[see eq.~(\ref{omegagw})]; $f$ is the frequency of the gravitational wave 
under consideration.

The generation of gravitational waves from violent bubble collisions
during a strongly first-order QCD transition has been suggested by 
Witten \cite{Witten}. The production of gravitational waves is 
possible if detonation bubbles collided, i.e.\ the bubble walls move
faster than the speed of sound. A more quantitative analysis for the 
collision of
vacuum bubbles (all latent heat goes into the kinetic energy of the 
bubble walls) has been performed by Kosovsky et al.~\cite{Kosovsky}.
However, neither in the bag model nor from lattice QCD it is likely
that a bubble nucleation scenario with detonation waves (almost vacuum 
bubbles) takes place.
In the opposite, the most likely scenario is the occurence of weakly 
deflagrating bubbles
\cite{Ignatius}. The generation of gravitational waves under these 
circumstances has been investigated by Hogan \cite{Hogangw} and Kamionkowski 
et al.~\cite{Kamionkowski}. An estimate in \cite{Hogangw} yields
$\Omega_{\rm gw} \sim 10^{-5} (d_{\rm nuc} H)^3 c_s^6$, 
for $f \sim H$ at the time of the QCD transition, which is $f\sim 10^{-7}$ Hz 
today. These gravitational waves are induced by the large inhomogeneities in 
energy density due to the coexistence of the QGP and the HG during a period
$\sim H^{-1}$. The most optimistic scenarios allow $\Omega_{\rm gw} \sim 
10^{-13}$; however, scenarios based on recent lattice QCD results give 
$\Omega_{\rm gw} \sim 10^{-23}$, which is completely out of reach of any 
technique known today for the detection of gravitational waves. The 
calculations in \cite{Kamionkowski} result in a different dependence on 
$d_{\rm nuc} H$; however, $\Omega_{\rm gw}$ is as small as from \cite{Hogangw}.
\smallskip 

\noindent 
Formation of black holes 

Crawford and Schramm \cite{Crawford} suggested that long-range forces during 
the QCD transition lead to the formation of planetary mass black holes.
However, there is no evidence that there are such large correlation lengths
at the QCD transition (km instead of fm!). 
Hall and Hsu \cite{Hall} argued that collapsing
bubbles might collapse to black holes. The mass of these objects would
be at most $\rho V_{\rm bubble}$, which, if the typical bubble scale is 
$1$ cm, is $< 10^{-18} M_{\odot}$. Since the supercooling in the cosmological
QCD transition is tiny, such a violent collapse of the walls
of the shrinking quark droplets is impossible. Moreover, black holes of 
these masses could not have survived until today, they would have evaporated 
long ago \cite{Carr}. 

The vanishing of the speed of sound during the coexistence phase also leads to
interesting gravitational effects, as pointed out by Jedamzik \cite{Jedamzik,
Jedamzik2}.
It was shown by several groups that the drop of the speed of sound is 
not sufficient to give rise to black hole formation \cite{SSW2,Cardall,SW}.

\section{The radiation fluid at the QCD scale}

The expansion of the Universe is very slow with respect to the strong,
electromagnetic, and weak interactions around $T_\star$ 
(see figure \ref{fig2}). To be more explicit,
the rate of the weak interactions is $\Gamma_{\rm w} \sim G_{\rm F}^2
T_\star^5 \approx$ $10^{-14}$~GeV, the rate of the electromagnetic interactions
is $\Gamma_{\rm em} \sim \alpha^2 T_\star \approx 10^{-5}$ GeV,
and the rate of the strong interactions is $\Gamma_{\rm s} \sim
\alpha^2_{\rm s}(T_\star) T_\star \approx 10^{-1}$ GeV. These rates have
to be compared with the Hubble rate $H \sim T_\star^2/m_{\rm Pl} \approx
10^{-21}$ GeV. Thus, leptons, photons, and the QGP/HG are in thermal and
chemical equilibrium at cosmological time scales. All components have
the same temperature locally, i.e.~smeared over scales $\lambda \sim
10^{-7} R_{\rm H}$. At scales $\lambda > 10^{-7} R_{\rm H}$, strongly, weakly,
and electromagnetically interacting matter makes up a single perfect
(i.e. dissipationless) radiation fluid.

There are no conserved quantum numbers for the radiation fluid, apart from
the lepton numbers and the baryon number. However, the corresponding 
chemical potentials are negligibly small at the QCD epoch. For the baryon 
chemical potential this is shown below (section \ref{baryons}). All lepton 
chemical 
potentials are assumed to vanish exactly. In this situation all thermodynamic
quantities follow from the free energy. The free energy density is 
$f(T) = - p(T)$ from homogeneity. The entropy density is given by the 
Maxwell relation for the free energy:
\begin{equation}
   \label{s}
   s = {{\rm d}p \over {\rm d}T} \ ,
\end{equation}
and the energy density $\epsilon(T)$ follows from the second law of 
thermodynamics for reversible processes:
\begin{equation}
  \label{2ndlaw}
  \epsilon = T {{\rm d} p \over {\rm d} T} - p \ .
\end{equation}

\subsection{Equation of state}

\begin{figure}[t]
\vspace*{0.3 cm}
\centerline{\includegraphics[width=0.65\textwidth]{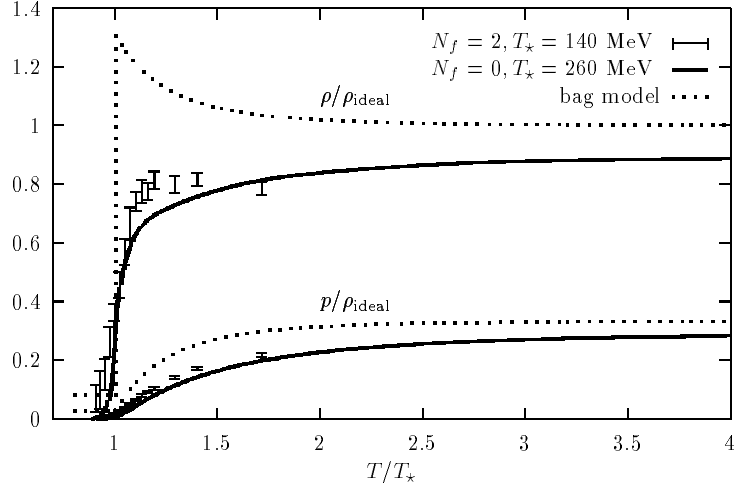}}
\caption{\label{fig4}
The energy density and the pressure of hot QCD, both relative to the 
energy density of a Stefan--Boltzmann (ideal) gas of quarks and gluons,
are plotted as functions of $T/T_\star$ (from \protect\cite{SSW2}). The 
lattice QCD data for $N_f = 0$ \protect\cite{Boyd} and $N_f = 2$ 
\protect\cite{MILC96} are compared with the predictions of the bag model.}
\end{figure}

The behaviour of $\epsilon(T)$ and $p(T)$ near the QCD transition must be 
given by non-perturbative methods, by lattice QCD. In figure \ref{fig4} 
lattice QCD data are shown for $\epsilon(T)$ (denoted by $\rho$ in the figure)
and $p(T)$ divided by $\epsilon$ 
of the corresponding SB gas. The lattice results for two 
systems are plotted: quenched QCD (no quarks)
\cite{Boyd}, and two-flavour QCD \cite{MILC96}. For quenched QCD the lattice
continuum limit is shown. For two-flavour QCD the data with six time steps
($N_t = 6$, $a\approx 0.2$ fm) and a quark mass $a m_q = 0.0125$ is
shown. This corresponds to a physical mass $m_q \sim 14$ MeV, a bit heavier
than the physical masses of the up and down quarks. On the horizontal axis we
plot $(T/T_{\star})$.  For $T/T_{\star} = 4$, 
energy density and pressure for quenched QCD are still 10\% resp. 15\% below 
the SB gas value. This is in excellent agreement with analytic calculations 
at finite temperature \cite{Blaizot99,Laine02}. 
It is remarkable that $\epsilon/\epsilon_{\rm SB}$ and 
$p/\epsilon_{\rm SB}$ versus $T/T_{\star}$ is quite similar for quenched 
QCD and two-flavour QCD.  Moreover, the temperature dependence of the 
rescaled pressure for four-flavour QCD \cite{Karsch97} is quite similar to 
quenched QCD. For more lattice QCD results see \cite{KarschRev}.
At temperatures below $T_{\star}$ quarks and gluons are confined to hadrons,
mostly pions. At present the hot pion phase is not seen in the two-flavour
lattice QCD, since the pion comes out too heavy ($0.3 <
m_\pi/m_\rho < 0.7$ from \cite{MILC96}, while the physical ratio is $0.18$).

In the bag model (see section \ref{bagm}) the pressure was given in 
eqs. (\ref{pqgp}) and (\ref{phg}) and thus from (\ref{s}) and (\ref{2ndlaw}) 
the energy density and entropy density read:
\begin{eqnarray}
\label{ener}
\epsilon_{\rm QGP} &=& \epsilon^{\rm SB}_{\rm QGP}+ B \\
\label{ent}
s_{\rm QGP} &=& s_{\rm QGP}^{\rm SB} \ ,
\end{eqnarray}
with $\epsilon^{\rm SB} = (\pi^2/30) g T^4$ and $s^{\rm SB} =
(2\pi^2/45) g T^3$ for massless particles.
Note that $s/T^3$ is a simple step function in the bag model.

\subsection{Adiabatic expansion \label{adex}}

Entropy is conserved, apart from the very short stage of reheating
($\sim 10^{-6} t_{\rm H}$) after the first bubbles have been nucleated.
This allows us to calculate $T=T(a)$ from ${\rm d}(s(T) a^3) = 0$, i.e.
\begin{equation}
\label{Ta}
{{\rm d}T\over{\rm d}\ln a} = - {3 s\over {\rm d} s/{\rm d} T} \ ,
\end{equation}
except for $T = T_\star$ in the case of a first-order phase transition.
In the bag model, $T \propto 1/a$ for $T\neq T_\star$.

The expansion while the QGP and HG  coexist in a first-order QCD transition 
is determined by entropy conservation,
\begin{equation}
a_+ = \left({s_-\over s_+}\right)^\frac13 a_- \ ,
\end{equation}
where the index $-(+)$ denotes the value of a quantity at the beginning 
(end) of the coexistence epoch. In the bag model the Universe expands by 
a factor of $a_+/a_- \approx 1.4$ until all QGP has been converted into the HG,
whereas for a lattice QCD fit \cite{SSW,SSW2} the Universe expands by a 
factor of $a_+/a_- \approx 1.1$. The growth of the scale factor is related 
to a lapse in cosmic time by ${\rm d}\ln a = {\rm d} t$. In terms of the 
Hubble time the transition lasts $0.3 t_{\rm H}$, resp. $0.1 t_{\rm H}$ for 
the bag model resp. the lattice QCD fit of \cite{SSW,SSW2}.

During a first-order QCD transition, i.e.~$T = T_\star$, the pressure 
$p(T_\star) \equiv p_\star$ is constant. For any first-order QCD phase 
transition the energy density $\rho(a)$ is obtained from the first law
of thermodynamics ${\rm d}\epsilon = -3(\epsilon + p_\star){\rm d}a/a$. 

\subsection{Speed of sound\label{sound}}

The speed of sound relates pressure gradients to density
gradients, i.e. $\nabla p = c_s^2 \nabla \epsilon$. This relation is 
essential for
the evolution of density fluctuations. During the short period of 
supercooling the relation between
temperature and time strongly depends on $c_s^2$. This is important for 
the correct estimate of the mean bubble nucleation distance.

When analysing cosmological perturbations during the QCD transition,
for wavelengths $\lambda > 10^{-4} R_{\rm H}$, neutrinos are 
tightly coupled: $\Gamma_{\nu}/k \gg 1$. For these 
wavelengths the radiation fluid behaves as a perfect (i.e. dissipationless) 
fluid, entropy in a comoving volume is conserved, and the process is thus
reversible.
On the other hand, below the neutrino diffusion scale, $\lambda < 10^{-4} R_H$,
acoustic oscillations are damped away before the QCD transition (see section 
\ref{nudamp}).

\begin{figure}[t]
\centerline{\includegraphics[width=0.65\textwidth]{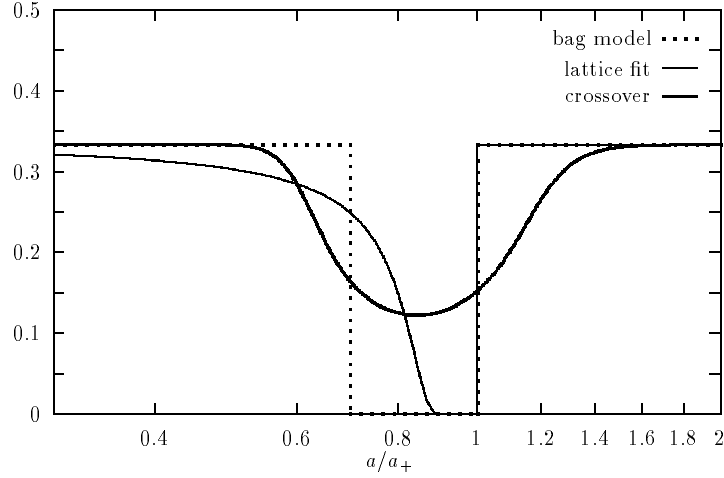}}
\caption{\label{fig5}
The behaviour of the speed of sound $c_s = (\partial p/
\partial\epsilon)_s^{1/2}$ during the QCD transition as a function of $a$ 
(from \protect\cite{SSW2}).
During a first-order transition (lattice fit and bag model) the speed
of sound vanishes.}
\end{figure}
The isentropic speed of sound (for wavelengths $\lambda$ much larger than 
the bubble separation), given by
\begin{equation}
\label{cs2}
c_s^2 = \left({\partial p\over \partial \epsilon}\right)_s =
{{\rm d} p/{\rm d}a\over {\rm d} \epsilon/{\rm d}a} \ ,
\end{equation}
must be zero during a first-order phase transition for a fluid with negligible
chemical potential (i.e. no relevant conserved quantum number), because 
$p(a) = p_\star =$ constant. 

In the bag model, $c_s^2 =1/3$ before and after the 
transition and vanishes during the transition\footnote{A small drop of 
the speed of sound was found by Dixit and Lodenquai \cite{Dixit} in a bag 
model taking interactions 
and masses into account. However, they missed the fact that the speed of 
sound drops to 
zero in a cosmological first-order QCD transition.}.
For a crossover the speed of sound decreases below $1\sqrt{3}$ at $T_\star$,
but does not drop to zero (see \cite{Blaizot} for a simple analytic model 
of a crossover transition).

The isentropic condition applies during the part of the phase transition 
after the initial supercooling, bubble nucleation, and sudden reheating 
to $T_{\star}$. During the adiabatic part of the transition, which
takes about $99\%$ of the transition time, the fluid is extremely close to
thermal equilibrium, because the time to reach equilibrium is very much
shorter than a Hubble time, i.e. the fluid makes a reversible
transformation. This can be seen as follows: across the bubble walls, local
pressure equilibrium is established immediately, $p_{\rm QGP} = p_{\rm HG}$
locally. The local temperature equilibrium, $T_{\rm QGP} = T_{\rm HG}$, is
established by neutrinos, which have a mean free path of $10^{-6} R_{\rm H}$,
enormously larger than the bubble wall thickness, and a collision time much
shorter than the Hubble time. This local pressure and temperature equilibrium
can only be satisfied if $p=p_{\star}$ and $T=T_{\star}$ at the bubble
walls. Over distance scales of the order of the bubble separation ($\sim 1$
cm), pressure (and therefore also temperature) is equalized with the velocity
of sound, and thereby the released latent heat is distributed. This pressure
equalization is very fast with respect to the Hubble expansion velocity 
$v_{\rm H} \approx 10^{-6} c$ at the $1$ cm scale. 

{}From eqs.~(\ref{2ndlaw}) and (\ref{s}) the speed of sound may be evaluated 
as 
\begin{equation}
\label{cs2s}
c_s^2 = ({\rm d}\ln s/{\rm d}\ln T)^{-1} \ . 
\end{equation}
The relation between temperature and time during the adiabatic expansion 
depends on the speed of sound. From eqs.~(\ref{Ta}) and (\ref{cs2s}) we find
\begin{equation}
{\rm d} \ln T  = - 3c_s^2 H {\rm d} t \ .
\end{equation}
This relation holds during the short period of supercooling before the 
first bubbles have been nucleated. An estimate of $c_s^2$ in the 
QGP at $T_\star$ has been obtained from lattice QCD.
 
A strong decrease in the speed of sound, already above $T_\star$, has been 
observed in lattice QCD for $N_f = 0$ \cite{Boyd} and for $N_f = 2$ 
\cite{MILC96,CP-PACS2flavour}. 
In both cases, $c_s^2(T_\star) \approx 0.1$ when approaching the critical
temperature from above (in the QGP). Note that this implies that the 
speed of sound relevant to bubble nucleation is of order $0.3$ instead of
$1/\sqrt{3} \approx 0.6$. 

\subsection{Baryons \label{baryons}}

The baryons are tightly coupled to the radiation fluid at the QCD scale.
Their energy density is negligible with respect to that of the
relativistic particles (photons, leptons, quarks/pions), thus they are 
dragged with the radiation fluid. Below I argue that the baryon chemical
potential is negligible at the QCD scale.

At high temperatures ($T > T_\star$) the baryon number density
may be defined as $n_{\rm B} \equiv \frac 13 \sum 
(n_{\rm q} - n_{\bar{\rm q}})$,
where $n_{\rm q}$ ($n_{\bar{\rm q}}$) is the number density of a specific quark
(antiquark) flavour, and the sum is taken over all quark flavours. At $T < 1$
GeV only the u, d, and s quarks contribute significantly. At low temperatures
($T < T_\star$) the baryon number density is defined as $n_{\rm B}
\equiv \sum (n_{\rm b} - n_{\bar{\rm b}})$; now the summation is taken over
all baryon species. Practically the nucleons contribute to the baryon number
of the Universe only.

Below the electroweak transition ($T_{\rm EW} = 100$--$200$ GeV) the baryon
number $B$ in a comoving volume is conserved. On the other hand, the 
entropy $S$ is conserved.
As a consequence the ratio of baryon number density and
entropy density $s$ is constant. From the abundances of primordial
${}^4$He and D we know the ratio $n_{\rm B}/n_{\gamma} =
(5.6 \pm 0.5) \times 10^{-10}$ \cite{bbn}.
Taking into account the three massless neutrinos along with the photons that
contribute to the entropy density, we find
\begin{equation}
\label{nB}
{n_{\rm B}\over s} = (7.5 \pm 0.7) \times 10^{-11} \ .
\end{equation}
Owing to the smallness of this ratio, the number of quarks equals the
number of antiquarks in the very early Universe.

Let me now turn to the baryon chemical potential. At high temperatures
the quark chemical potentials $\mu_{\rm q}$ are equal, because weak
interactions keep them in chemical equilibrium (e.g.\ u + e $\leftrightarrow$
d or s + $\nu_{\rm e}$), and the chemical potentials for the leptons are
assumed to vanish (see \cite{WeinbergBook} for a discussion of lepton chemical
potentials). Thus, the chemical potential for a baryon is defined by
$\mu_{\rm B} \equiv 3 \mu_{\rm q}$. For an antibaryon the chemical potential
is $ - \mu_{\rm B}$. The baryon number density of an SB Fermi gas
of three quark flavours reads $n_{\rm B}
\approx T^2\mu_{\rm B}/3$ at high $T$. From eq.~(\ref{nB}) one
finds that
\begin{equation}
{\mu_{\rm B}\over T} \sim 10^{-9} \qquad \mbox{at\ }
T > T_\star \ .
\end{equation}

At low temperatures ($T< T_\star$), $\mu_{\rm B} = \mu_{\rm p} =
\mu_{\rm n}$, neglecting the mass difference between the proton and the
neutron. The ratio of baryon number density and entropy now reads
\begin{equation}
\label{mu}
{n_{\rm B}\over s} \approx 0.05 \left({m_{\rm p}\over T}\right)^{\frac{3}{2}}
\exp\left(-{m_{\rm p}\over T}\right) \sinh\left({\mu_{\rm B}\over T}\right) \ .
\end{equation}
Since this ratio is constant, the behaviour of $\mu_{\rm B}/T$ is given by
eq.\ (\ref{mu}), e.g.\ at $m_{\rm p}/T \approx 20$ one finds
$\mu_{\rm B}/T \approx 10^{-2}$. All antibaryons
are annihilated when the ratio
\begin{equation}
{n_{\rm b} - n_{\bar{\rm b}}\over n_{\rm b} + n_{\bar{\rm b}}} \approx
\tanh\left({\mu_{\rm B}\over T}\right)
\end{equation}
goes to unity. This happens when $\mu_{\rm B}/T \sim 1$, which corresponds to
$T_{\rm ann} \approx m_{\rm p}/25 \approx 40$ MeV.
Below this temperature the baryon chemical
potential is $\mu_{\rm B}(T\ll T_{\rm ann}) \approx m_{\rm p}$.
To add one proton to the Universe one proton rest mass should be invested.
A detailed investigation of the baryon chemical potential in the early Universe
was recently given by \cite{Rafelski}. 

It was argued above that, from sphaleron processes before the electroweak 
transition, a possible lepton asymmetry should be of the same order of 
magnitude as the baryon asymmetry. Thus, at temperatures $T \gg m_e$ 
we find that lepton chemical potentials are negligible, since $\mu_{\rm L}/T 
\sim n_{\rm L}/T^3 \sim n_{\rm B}/T^3 \sim \eta$. 
This justifies the previous assumption that any chemical potential can be 
neglected during the first second. 

\section{Evolution of gravitational waves}

In principle, primordial gravitational waves (e.g. from cosmological inflation)
present a very clean probe of the dynamics of the early Universe, since 
they know only about the Hubble expansion. As was shown in \cite{Schwarz}
a step is imprinted in the spectrum of primordial gravitational waves  
by the cosmological QCD transition. This step does not allow us to tell 
the difference between a first-order transition and a crossover, 
but its position would allow an estimate of the temperature and its height 
would allow a measurement of the change in the effective number of relativistic 
degrees of freedom.

Primordial gravitational waves are predicted to be generated during 
inflation \cite{Starobinskii,Abbott} and could be detected by observing the
so-called B-mode (parity odd patterns) polarisation of the CMB. 
Inflation predicts an almost scale-invariant energy density per 
logarithmic frequency 
interval for the most interesting frequencies ($\sim 10^{-8}$ Hz for pulsar 
timing, $\sim 10^{-3}$ Hz for LISA, and $\sim 100$ Hz for LIGO and VIRGO) of 
the gravitational waves.

In the cosmological context, the line element of gravitational waves is 
given by
\begin{equation}
{\rm d}s^2 = -{\rm d}t^2 + a^2(\delta_{ij}+h_{ij}){\rm d}x^i{\rm d}x^j \ ,
\end{equation}
where $h_{ij}$ is a transverse, traceless tensor. The linearised Einstein
equation admits wavelike solutions for $h_{ij}$\footnote{For a more general 
definition and discussion of gravitational waves see Ref.~\cite{ZBook}.}. 
The spatial average 
\begin{equation}
\langle h_{ij}(x)h^{ij}(x+r) \rangle = \int j_0(kr) k^3 |h_k|^2 {\rm d}\ln k
\end{equation}
defines the power spectrum $|h_k|^2$. We denote by $h$ the rms amplitude
of a gravitational wave per logarithmic frequency interval:
$h \equiv k^{3/2}|h_k|$. The linearized equation of motion for $h(t)$ reads
\begin{equation}
\label{ev}
\ddot{h} + 3 H \dot{h} + {k^2\over a^2} h = 0 \ ,
\end{equation}
where the differentiation is taken with respect to cosmic time $t$.
The amplitude of gravitational waves is constant on superhorizon scales
and decays as $1/a$ after horizon crossing,
$h \simeq C_k \sin(k\eta + \delta_k)/a $,
where $\eta = \int {\rm d}t/a$ is conformal time;
$C_k$ and $\delta_k$ are determined by matching the subhorizon 
to the superhorizon solution.

For subhorizon modes, $k_{\rm ph} \gg H$, the energy density of gravitational 
waves can be defined. The space-time average of the energy-momentum tensor 
over several wavelengths gives $\epsilon_{\rm gw} = - (1/32 \pi G)
\langle \dot{h}_{ij} \dot{h}^{ij}\rangle$. The energy density per logarithmic
interval in $k$ is related to the rms amplitude $h$:
\begin{equation}
\label{rhodef}
k{{\rm d}\epsilon_{\rm gw}\over {\rm d}k} = {1 \over 32 \pi G}
k_{\rm ph}^2 \frac 12 h^2 \ .
\end{equation}
The factor $1/2$ comes from the time average over several oscillations.
The energy fraction in gravitational waves, per logarithmic interval in $k$,
is defined by
\begin{equation}
\label{omegagw}
\Omega_{\rm gw} (k) \equiv  k {{\rm d} \epsilon_{\rm gw}\over {\rm d} k}
{1\over \epsilon_{\rm c}} \ ,
\end{equation}
where $\epsilon_c \equiv 3 H_0^2/8\pi G$. 

\begin{figure}
\begin{center}
\input{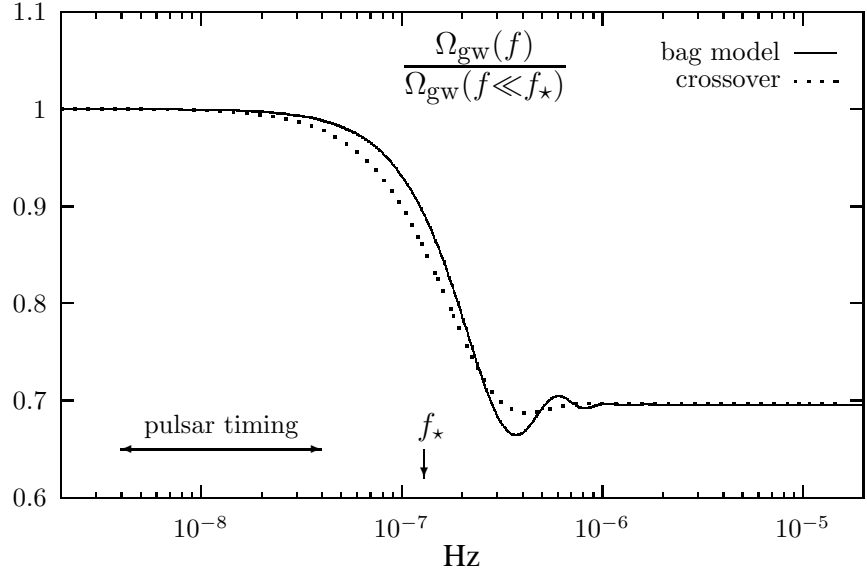}
\end{center}
\caption{The modification of the energy density, per logarithmic frequency
interval, for primordial gravitational waves from the QCD transition 
(from \protect\cite{Schwarz}).
\label{fig6}}
\end{figure}

Figure \ref{fig6} shows the transfer function $\Omega_{\rm gw}(f)/
\Omega_{\rm gw}(f \ll f_\star)$ from a numerical integration of eq.~(\ref{ev})
through the cosmological QCD transition. The typical frequency scale is
\begin{equation}
f_\star \approx 1.36 \left(g\over 17.25\right)^{\frac12}
{T_\star \over 150 \mbox{\ MeV}} 10^{-7} \mbox{\ Hz} \ ,
\end{equation}
which corresponds to the mode that crosses the Hubble horizon at the end of
the bag model QCD transition. Scales that cross into the horizon after the
transition (l.h.s. of the figure) are unaffected, whereas modes that
cross the horizon before the transition are damped by an additional factor
$\approx 0.7$. The modification of the differential spectrum has been
calculated for a first-order (bag model) and a crossover QCD transition.
In both cases the step extends over one decade in frequency. The detailed
form of the step is almost independent from the order of the transition.

The size of the step can be calculated analytically \cite{Schwarz}.
Comparing the differential energy spectrum for modes that cross into the 
horizon before and after the transition gives the ratio
\begin{equation}
\label{ratio}
{\Omega_{\rm gw}(f \gg f_\star) \over \Omega_{\rm gw}(f\ll f_\star)} =
\left(g_{\rm a}\over g_{\rm b}\right)^{\frac13} \approx 0.696 \ ,
\end{equation}
for the QCD transition, which coincides with the numerical integration in
fig.~\ref{fig6}. This result is in agreement with the entropy conservation
of subhorizon gravitational waves \cite{Hogangw,Krauss,Kolb90}. However, 
for superhorizon modes the entropy is not defined.

Similar steps in the differential spectrum have been studied for
gravitational waves generated by cosmic strings \cite{Vilenkin}.
These gravitational waves are generated on subhorizon scales
when the cosmic strings decay. Their frequency is larger than $\alpha^{-1} H$,
where $\alpha$ is the ratio between the typical size of a string loop
and the Hubble radius at formation of the loop; $\alpha$ is smaller than
$0.1$ and might be as small as $10^{-5}$ \cite{Caldwell}.
In this situation, steps in the differential spectrum follow from the
conservation of entropy for decoupled species (gravitons) during cosmological
phase transitions \cite{Bennett}.
This interpretation applies to modes that have been inside the horizon
long before the transition or that have been generated on subhorizon scales.
However, for superhorizon modes,
entropy and energy of a gravitational wave are not defined.
We therefore cannot rely on a conservation-of-entropy argument when
dealing with gravitational waves from inflation.

In fig.~\ref{fig6} we indicated the frequency range ($\sim 1$ yr$^{-1}$)
in which limits on $\Omega_{\rm gw}$ have been reported
from pulsar timing residuals \cite{pulsar}.
The frequencies where the step of the QCD transition would be visible is of
the order of $0.3$ month$^{-1}$. For pulsar timing, the power spectrum of
gravitational waves is more relevant than the energy spectrum.
The power spectrum is $\propto \Omega_{\rm gw}(f) f^{-5}$. Our results show
that the power spectrum might deviate from the $f^{-5}$ behaviour
over a whole decade in frequency. Unfortunately, with today's technology 
we will not be able to detect primordial gravitational waves at frequencies
around $10^{-7}$ Hz, because their amplitude is expected to be to small.

\section{A first-order QCD transition \label{bubbles}}

In a first-order QCD transition the quark--gluon plasma supercools 
before the first bubbles of hadron gas are formed. In a homogeneous Universe
without `dirt' the bubbles nucleate owing to thermal fluctuations 
(homogeneous nucleation). 
 
If cosmic `dirt' in the form of defects (such as strings) or black holes is
present at the QCD epoch, this `dirt' may trigger the formation of 
the first hadronic bubbles (like the nucleation of vapour bubbles in a pot 
of boiling water). 

There is a broad range in parameter space, where the magnitude of
primordial temperature fluctuations is of the same order or larger than 
the typical supercooling in the homogeneous nucleation scenario. In this 
case, the transition proceeds with inhomogeneous bubble nucleation. The 
mean nucleation distance results from the scale and amplitude of the 
temperature fluctuations. 

\subsection{Homogeneous nucleation \label{homn}}

The probability to nucleate a bubble by a thermal fluctuation is 
proportional to $\exp(\Delta S)$, where $\Delta S$ is the change in entropy
by creation of a hadronic bubble. The second law relates $\Delta S$ to the 
minimal
work done in this process, which is the change in the free energy because
the volume and temperature are fixed. The change in free energy 
of the system by creating a spherical bubble with radius $R$ is
\begin{equation}
\Delta F = {4\pi\over 3}(p_{\rm QGP} - p_{\rm HG})R^3 + 4\pi \sigma R^2 \ ,
\end{equation}
where $\sigma$ is the surface tension. Bubbles can grow if they are
created with radii greater than the critical bubble radius $R_{\rm c}$.
Smaller bubbles disappear again, because the free energy gained from the
bulk of the bubble is more than compensated by the surface energy in the 
bubble wall; $R_{\rm c}$ is determined from the maximum value of 
$\Delta F(R)$ and reads
\begin{equation}
R_{\rm c}(T) = {2 \sigma \over p_{\rm HG}(T) - p_{\rm QGP}(T)} \ .
\end{equation}
At $T_\star$ the critical bubble size diverges, and no bubble can be formed.
Finally, the probability to form a hadronic bubble with critical radius
per unit volume and unit time is given by
\begin{equation}
\label{I}
I(T) = I_0(T) \exp\left( - {\Delta F_{\rm c}\over T}\right) \ ,
\end{equation}
with $\Delta F_{\rm c} = 16\pi\sigma^3/[3(p_{\rm HG} - p_{\rm QGP})^2]$.
For dimensional reasons the prefactor $I_0 \sim C T_\star^4$, with 
$C = {\cal O}(1)$. A more detailed calculation of $I_0$ within the bag model 
has been provided in \cite{Csernai}.
It was shown in Ref.~\cite{Christiansen} that the temperature dependence
of the prefactor $I_0$ can be neglected for the calculation of the
supercooling temperature $T_{\rm sc}$ in the cosmological QCD
transition. As will be shown below, the numerical prefactor $C$ is 
irrelevant in the cosmological QCD transition.

For small supercooling $\Delta \equiv 1 -  T/T_\star \ll 1$ we may evaluate
$(p_{\rm HG} - p_{\rm QGP})(T)$ by using the second law of thermodynamics,
i.e.\ $p_{\rm HG} - p_{\rm QGP} \approx l \Delta$, and thus
\begin{equation}
\label{A}
I(\Delta) \approx I_0(T_\star) \exp\left(-A/\Delta^2\right),
\end{equation}
with $A \equiv 16 \pi \sigma^3/(3 l^2 T_\star)$ and $I_0(T_\star) \approx
T_\star^4$. Note that this result does not depend on the details of the 
QCD equation of state. For the values of $l=1.4 T_\star^4$ and 
$\sigma=0.015 T_\star^3$ from quenched lattice QCD \cite{Iwasaki2} 
$A \approx 3 \times 10^{-5}$. In the bag model $A \approx 5 \times 10^{-2} 
(\sigma/T_\star^3)^3$. 

The amount of supercooling that is necessary to complete the transition,
$\Delta_{\rm sc}$, can be estimated from the schematic case of one single 
bubble nucleated per Hubble volume per Hubble time, which is 
\begin{equation}
\label{Delta_sc}
{\cal O}(\Delta_{\rm sc}) = \left[\frac{A}{4 \ln(T_\star/H_\star)} \right]^{1/2}
\approx 4 \times 10^{-4} 
\end{equation}
for the values of $l$ and $\sigma$ from quenched lattice QCD \cite{Iwasaki2}. 
For the bag model I assume  $\sigma < 0.1 T_\star^3$, which implies that 
$\Delta_{\rm sc} < 6 \times 10^{-4}$. 
It has been shown in \cite{Ignatius} that for such a tiny supercooling
the formation of detonation bubbles is forbidden. This justifies the 
approximation of small supercooling made above.  

The time lapse during the supercooling period follows from the conservation
of entropy and reads
\begin{equation}   
\Delta t_{\rm sc}/t_{\rm H} = \Delta_{\rm sc}/(3 c_s^2) = {\cal O}(10^{-3}) \ . 
\end{equation}
Here we used the relation $c_s^2 = {\rm d} \ln s/{\rm d} \ln T$ for the
speed of sound in the supercooled phase. For realistic models 
$0 < c_s(\Delta) < 1/\sqrt{3}$. In the bag model $c_s(\Delta) = 1/\sqrt{3}$.

The critical size of the bubbles created at the supercooling temperature is
\begin{equation}
R_{\rm c}(\Delta_{\rm sc}) \approx {2\sigma\over l \Delta_{\rm sc}} \approx
               25 \mbox{ fm} \ .
\end{equation}
The critical radius is large with respect to the QCD scale, and this 
justifies the thin wall approximation, which was made implicitly above.

\begin{figure}[t]
\centerline{\includegraphics[width=0.55\textwidth]{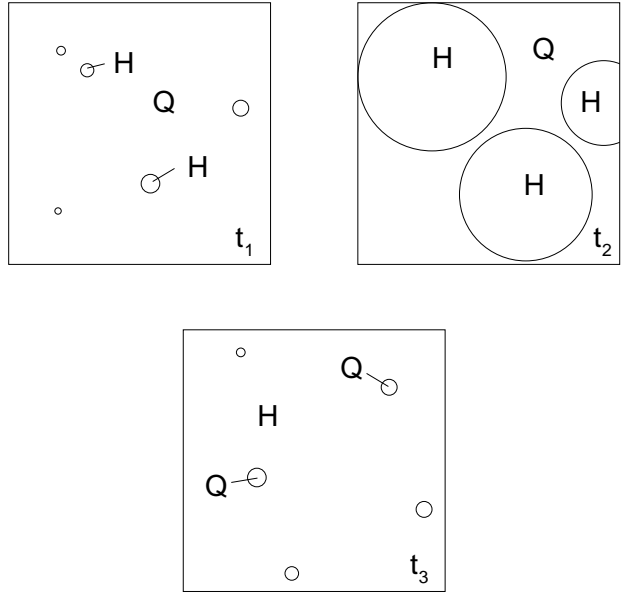}}
\caption{\label{fig7}
Sketch of a first-order QCD transition via homogeneous bubble nucleation: 
above the critical temperature the Universe is filled with a quark--gluon 
plasma (Q). After a small amount of supercooling the first 
hadronic bubbles (H) nucleate at $t_1$, with mean separation $d_{\rm nuc}$. 
At $t_2 > t_1$ these bubbles have grown and have released enough 
latent heat to quench the formation of new bubbles. The supercooling, 
bubble nucleation, and  quenching takes just $1\%$ of the full transition time.
In the remaining $99\%$ of the transition time the bubbles grow following  
the adiabatic expansion of the Universe. At $t_3$ the transition is almost 
finished. The shrinking QGP drops are separated by the typical distance 
$d_{\rm nuc}$.}
\end{figure}

After the first bubbles have been nucleated, they grow most probably
by weak deflagration \cite{DeGrand,Kurki-Suonio,Kajantie,Ignatius}.
The deflagration front (the bubble wall) moves with the velocity 
$v_{\rm defl} \ll 1/\sqrt{3}$ \cite{Kajantie92}. The energy that is 
released from the bubbles is distributed into the surrounding QGP by a 
supersonic shock wave and by neutrino radiation. This reheats the QGP 
to $T_\star$ and prohibits further bubble formation. Since the amplitude of 
the shock is very small \cite{Kurki-Suonio}, on scales smaller than the 
neutrino mean free path, heat transport by neutrinos is the most efficient. 
Neutrinos have a mean free path of $10^{-6} R_{\rm H}$ at $T_\star$. 
When they do most of the heat transport, heat goes with 
$v_{\rm heat}= {\cal O}(c)$. For larger scales, heat transport is much slower. 
Figure \ref{fig7} shows a sketch of the homogeneous bubble nucleation 
scenario. 

Let us now calculate the mean bubble separation, $d_{\rm nuc}$, and the 
final supercooling, $\Delta_{\rm sc}$, for a scenario with weak deflagration.
Bubbles present at a given time have typically been nucleated during the
preceding time interval 
\begin{equation}
\label{tnucl}
\Delta t_{\rm nuc} \equiv  I/({\rm d}I/{\rm d} t) \ .
\end{equation}
Using the relation between time and supercooling,
${\rm d}\Delta/{\rm d}t = 3 c_s^2 /t_{\rm H}$, we find
\begin{equation}
\Delta t_{\rm nuc}/t_{\rm H}= \Delta_{\rm sc}^3/(6 A c_s^2) = {\cal O}(10^{-5}) 
\end{equation}
and 
\begin{equation}
\label{Delta-nucl}
\Delta_{\rm nuc} = {\Delta t_{\rm nuc}\over \Delta t_{\rm sc}} 
\Delta_{\rm sc} = {\cal O}(10^{-2}) \Delta_{\rm sc} \ .
\end{equation} 
During the time interval $\Delta t_{\rm nuc}$ each bubble
releases latent heat, which is distributed over a typical  
distance $\approx \Delta 2 v_{\rm heat} t_{\rm nuc}$. This distance 
has a weak dependence on the precise value of $\Delta_{\rm sc}$, but the 
bubble nucleation rate increases strongly with $\Delta$ until 
one bubble per volume $\sim (\Delta t_{\rm nuc} v_{\rm heat})^3$ is 
nucleated. Therefore the mean bubble separation is
\begin{equation}
\label{dnucl}
d_{\rm nuc} \approx 2 v_{\rm heat} \Delta t_{\rm nuc} \approx 
\frac{v_{\rm heat}}{3 c_s^2}  {\Delta_{\rm sc}^3\over A}  R_{\rm H} = 
{\cal O}(10^{-6} R_{\rm H})= {\cal O}(1 {\rm cm}),
\end{equation}
where I used $v_{\rm heat} = {\cal O}(0.1), 3c_s^2 = {\cal O}(0.1)$, 
which gives a typical value for the nucleation distance. The suppression of 
bubble nucleation due to already existing bubbles is neglected. 

The estimate (\ref{dnucl}) of the mean bubble separation applies if the 
released latent heat by means of sound waves and by neutrino free streaming is
sufficient to reheat the QGP to $T_\star$, i.e. to quench the nucleation of 
new bubbles. On the other hand the typical bubble separation could be
given by the rate of release of latent heat, i.e. by the bubble wall velocity
$v_{\rm defl}$. Since the period of supercooling lasts about $1 \% $ of the
time needed for completing the entire first-order phase transition, $1 \% $ of
the QGP must be converted to HG in the process of sudden reheating to
$T_{\star}$; the bubble radius at quenching must therefore reach $0.2$ of the
bubble separation, $R_{\rm bubble}\approx 0.2 d_{\rm nuc}$.  With $R_{\rm
bubble} \approx v_{\rm defl} \Delta t_{\rm nuc}$, and using the above relation
$d_{\rm nuc} \approx 2 v_{\rm heat} \Delta t_{\rm nuc}$, we require $v_{\rm
defl} \ge 0.4 v_{\rm heat}$ for consistency. If $v_{\rm defl}$ is smaller than
this, the limiting factor for quenching is the rate of
release of latent heat by bubble growth, and the bubble separation is
\begin{equation}
d_{\rm nuc} \approx 2 v_{\rm defl} \Delta t_{\rm nuc} \approx 
\frac{v_{\rm defl}}{3 c_s^2}  {\Delta_{\rm sc}^3\over A} R_{\rm H} \ ,
\end{equation}
i.e. the bubble separation will be smaller than the estimate
in eq.~(\ref{dnucl}).

We are now in a position to improve the estimate of $\Delta_{\rm sc}$:
one bubble nucleates in the volume $(v_{\rm heat}\Delta t_{\rm nuc})^3$ 
during $\Delta t_{\rm nuc}$. 
This can be written as
\begin{equation}
\label{Delta-exact}
1 \approx  (v_{\rm heat} \Delta t_{\rm nuc})^3 \Delta t_{\rm nuc} 
I(t_{\rm sc}) \ ,
\end{equation}  
which in terms of the supercooling parameter $\Delta_{\rm sc}$ is given by:
\begin{eqnarray}
1 &\approx & {v_{\rm heat}^3\over (3 c_s^2 A)^4} 
\left(T_\star\over H_\star\right)^4
\Delta_{\rm sc}^{12} \exp\left({- A\over\Delta_{\rm sc}^2}\right) \nonumber \\ 
&\approx &
10^{94} \Delta_{\rm sc}^{12} 
\exp\left({-2.89 \times 10^{-5}\over\Delta_{\rm sc}^2}\right) \ . 
\end{eqnarray}
Also the pre-exponential factor is smaller by a factor of $10^{20}$ than
the naive estimate (\ref{Delta_sc}), 
the amount of supercooling is just $20\%$ larger than in (\ref{Delta_sc}),
i.e.\ $\Delta_{\rm sc} = 5 \times 10^{-4}$. This also demonstrates that
numerical prefactors in (\ref{Delta-exact}) are irrelevant in the calculation
of $\Delta_{\rm sc}$. 

To summarize, the scales on which non-equilibrium phenomena occur are
given by the mean bubble separation, which is about $10^{-6} R_{\rm H}$.
The entropy production is tiny, i.e. $\Delta S/ S \sim 10^{-6}$, since the
supercooling is small $\sim 10^{-3}$. After supercooling, which lasts $10^{-3}
t_{\rm H}$, the Universe reheats in $\Delta t_{\rm nuc} \approx 10^{-6}
t_{\rm H}$. After reheating, the thermodynamic variables follow their
equilibrium values and bubbles grow only because of the expansion of the 
Universe.

\subsection{Heterogeneous nucleation \label{hetn}}

In the first-order phase transitions that we know from our everyday experience,
for example the condensation of water drops in clouds, the drops, i.e. the 
bubbles are nucleated at impurities (`dirt'). This could happen in the 
early Universe as well. Candidates for cosmic `dirt' are primordial
black holes, monopoles, strings, and other kinds of defects. Of course,
the existence of any of these objects has not been verified so far.
Nevertheless, let me discuss in what manner cosmic `dirt' would change 
the nucleation of bubbles. The following considerations are based on the 
work of Christiansen and Madsen \cite{Christiansen}.

Let $n$ be the number density of the impurities. Further assume that at 
time $t_{\rm i}$ with $t_\star \leq t_{\rm i} < t_{\rm sc}$ bubbles 
nucleate at the locations of the impurities. (It is easy to see that 
$t_{\rm i}$ is restricted to the mentioned interval: before $t_\star$ a bubble 
cannot grow because $T>T_\star$, after $t_{\rm sc}$ homogeneous nucleation 
already happened.) 

There are two limiting cases: if $n \gg d_{\rm nuc,\ hom}^{-3}$, 
the mean nucleation distance is $d_{\rm nuc} = n^{-1/3} \ll 
d_{\rm nuc,\ hom}$. If $n \ll d_{\rm nuc,\ hom}^{-3}$, the impurities are
so rare that $d_{\rm nuc} = d_{\rm nuc,\ hom}$. 

The most interesting situation occurs when the typical distance between the 
impurities is something bigger than the mean homogeneous nucleation distance.
But it should be small enough for the bubbles nucleated at the 
impurities to reheat the Universe just before homogeneous nucleation starts. 
For a quantitative estimate, let me determine the amount of supercooling for
heterogeneous nucleation. As in homogeneous nucleation [see
eq.\ (\ref{Delta-exact})] I use the condition that one bubble is nucleated per 
reheated volume, i.e.\ the sum of the probabilities to form a bubble from
an impurity and from a thermal fluctuation: 
\begin{equation}
\label{het1}
1 \approx n (v_{\rm heat} \Delta t)^3 + (v_{\rm heat} \Delta t_{\rm nuc})^3 
\Delta t_{\rm nuc} I(t_{\rm sc}) \ .
\end{equation}
$\Delta t = t_{\rm sc} - t_{\rm i}$ is the growth time of bubbles
from impurities and $\Delta t_{\rm nuc}$ the growth time of bubbles from
thermal fluctuations, as defined in (\ref{tnucl}). Note that here $t_{\rm sc}$
may be smaller than in homogeneous nucleation. If 
$\Delta t/\Delta t_{\rm nuc}$ is of order unity both mechanisms work 
independently and 
\begin{equation}
d_{\rm nuc} \approx {1\over (n + \Delta t_{\rm nuc} I(t_{\rm sc}))^{1/3}}
\leq d_{\rm nuc,\ hom} \ .
\end{equation}
The most interesting situation arises when $\Delta t/\Delta t_{\rm nuc} \gg 1$.
This means that the probability for  a bubble from a thermal
fluctuation is reduced because the available volume for thermal fluctuations
in the QGP has been reduced by a factor of $(1 - n (v_{\rm heat} \Delta t)^3)$. 
This factor should be taken into account in (\ref{het1}) which now reads
\begin{equation}
\label{het2}
1 \approx n (v_{\rm heat} \Delta t)^3 + (v_{\rm heat} \Delta t_{\rm nuc})^3
\Delta t_{\rm nuc} I(t_{\rm sc})[1 - n (v_{\rm heat} \Delta t)^3] \ .
\end{equation}
The unique solution to this equation is $n (v_{\rm heat} \Delta t)^3 \approx 1$
and thus
\begin{equation}
d_{\rm nuc} \approx {1\over \{n + \Delta t_{\rm nuc} I(t_{\rm sc})
[1 - n (v_{\rm heat} \Delta t)^3]\}^{1/3}}
\approx  n^{-1/3} \ .
\end{equation}
The maximal mean nucleation distance from heterogeneous nucleation is found 
from $\Delta t \leq \Delta t_{\rm sc,\ het}$ to be
\begin{equation}
\mbox{max} (d_{\rm nuc}) \approx 
v_{\rm heat} \mbox{max}(\Delta t_{\rm sc,\ het}) \approx
v_{\rm heat} \Delta t_{\rm sc,\ hom} \approx 10^{-3} R_{\rm H} = {\cal O}
(10 {\rm\ m}) \ .
\end{equation}
There, for a narrow range in parameter space, namely 
$n (v_{\rm heat} \Delta t)^3 \approx 1$, nucleation distances may be 
larger than in homogeneous nucleation by a factor of $100$, which means that
$n$ should be $n\sim 10^9/R_{\rm H}^3$.

As an example for `dirt' I consider primordial black holes (PBHs):
with the appropriate density, $n_{\rm PBH}(T_\star)\sim 10^9/R_{\rm H}^3$,
PBHs are produced at the temperature $T_{\rm PBH} \sim 10^7$ GeV [see 
eq.\ (\ref{nPBH})] with a mass of $m_{\rm PBH} \sim 10^{-16} M_\odot$.
PBHs of this mass are excluded observationally because 
they start to evaporate today and should be observed as $\gamma$-sources 
\cite{Carr}. Thus PBHs cannot be considered as seeds that yield the maximum
nucleation distance. Nevertheless, larger PBH masses still may seed bubbles,
but their mean separation is so large that thermal bubble nucleations cannot be
suppressed. 

Another example of `dirt' are cosmic strings. Recently, such a scenario has 
been analysed in \cite{Layek}. A moving cosmic string would generate an 
overdense plane, on which the phase transition would be delayed.  

\subsection{Inhomogeneous nucleation\label{ihn}}

The local temperature $T(t,{\bf x})$ of the radiation fluid fluctuates,
because cosmological perturbations have been generated during cosmological
inflation
\cite{Mukhanov}. Let me denote the temperature fluctuation by $\Delta_T
\equiv \delta T/T$. Inflation predicts a Gaussian distribution of 
perturbations:
\begin{equation}
P(\Delta_T){\rm d}\Delta_T = {1\over \sqrt{2\pi}\Delta_T^{\rm rms}}
\exp\left( - \frac12 {\Delta_T^2\over (\Delta_T^{\rm rms})^2}\right)
{\rm d}\Delta_T \ .
\end{equation}
If one allows for a tilt in the power spectrum
of density fluctuations, the COBE \cite{COBE} normalized rms temperature
fluctuation reads \cite{IS}
\begin{equation}
\Delta_T^{\rm rms} \approx 10^{-4} (3 c_s^2)^{3/4} 
\left({k\over k_0}\right)^{(n-1)/2} \ ,
\end{equation}
where $k_0$ is the wave number of the mode that crosses the Hubble radius today.
The case $n=1$ gives the Harrison--Zel'dovich spectrum \cite{HZ}. Recent 
WMAP results combined with other CMB data and 2dF galaxy redshift survey 
results give $n-1 = -0.03 \pm 0.03$ ($n-1 = - 0.07^{+0.04}_{-0.05}$) 
without (with) running of the spectral index (${\rm d}n/{\rm d}\ln k \neq 0$)
\cite{Spergel}. For $n = 1$ we find $\Delta_T^{\rm rms}(k_{\rm QCD}) 
\approx 2 \times 10^{-5}$.

{}From the above we conclude that $\Delta_T^{\rm rms}$ and $\Delta_{\rm nuc}$
[see eq.\ (\ref{Delta-nucl})]
may be of the same magnitude or that $\Delta_T^{\rm rms}$ may be even larger.
The picture of homogeneous bubble nucleation, where bubbles form from 
statistical fluctuations, is false for the most probable cosmological 
scenarios. 

\begin{figure}[t]
\centerline{\includegraphics[width=0.55\textwidth]{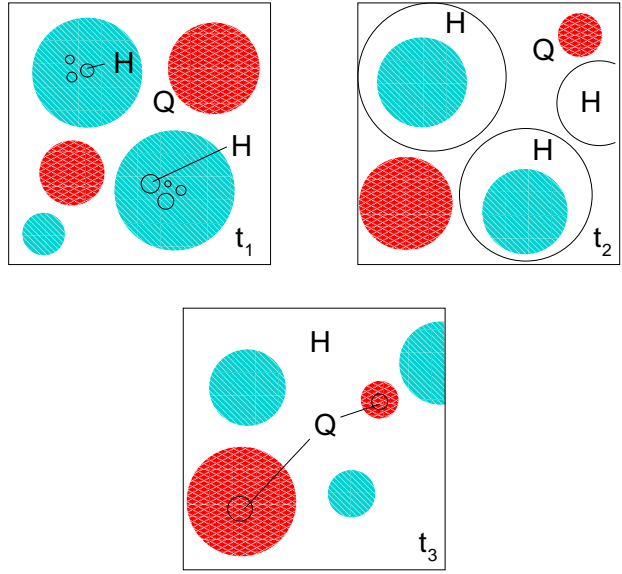}}
\caption{\label{fig8}
Sketch of a first-order QCD transition in the inhomogeneous
Universe (from \protect\cite{IS}): 
at $t_1$ the coldest spots (dark grey) are cold enough to render the
nucleation of hadronic bubbles (H) possible, while most of the
Universe remains in the quark--gluon phase (Q). At $t_2 > t_1$ the bubbles
from the cold spots have merged and have grown to bubbles as large
as the fluctuation scale. Only the hot spots (light grey) are still in the 
QGP phase. At $t_3$ the transition is almost finished. The last 
QGP drops are found in the hottest spots of the Universe. The mean 
separation of these hot spots can be much larger than the homogeneous 
bubble nucleation separation.}
\end{figure}
We thus developed a new scenario for the cosmological QCD transition
\cite{IS}. To do that we had to learn more about the primordial 
temperature fluctuations first.  
A small scale cut-off in the spectrum of primordial temperature fluctuations
comes from collisional damping by neutrinos \cite{Weinberg,SSW2} (see also
section \ref{nudamp}).
Because the neutrinos interact only weakly, their mean free path is large
with respect to the strong and electromagnetic interacting particles.
The interaction rate of neutrinos is $\Gamma \sim G_{\rm F}^2 T^5$.
This has to be compared with the frequency $c_s k_{\rm ph}$. We find that
at the QCD transition neutrinos travel freely on scales
$\lambda_{\nu-{\rm mfp}} \approx 10^{-6} R_{\rm H}$.
Fluctuations on the diffusion scale of neutrinos are washed out by the
time of the QCD transition (see section \ref{nudamp}):
\begin{equation}
\lambda_{\nu-{\rm diff}} =
\frac13 \sqrt{\lambda_{\nu-{\rm mfp}} c t_{\rm H}}
\approx 10^{-4} R_{\rm H} \ .
\end{equation}
Thus the old picture of homogeneous bubble nucleation still applies
within the small homogeneous patches of $\lambda_{\rm smooth} = 
10^{-4} R_{\rm H}$.

The compression time scale for a homogeneous patch is $\delta t =
\lambda_{\rm smooth}/c_s$ $\sim 10^{-3} t_{\rm H}$. If the compression
time scale is larger than $\Delta t_{\rm nuc}$ the temperature fluctuations are
frozen with respect to the time scale of nucleations.

A sketch of inhomogeneous bubble nucleation is shown in fig.~\ref{fig8}. The
basic idea is that temperature inhomogeneities determine the location of
bubble nucleation. In cold regions, bubbles nucleate first.
In general we have two possible situations:
\begin{enumerate}
\item If $\Delta_{\rm nuc} > \Delta_T^{\rm rms}$, the temperature
      inhomogeneities are negligible and the phase transition proceeds 
      via homogeneous nucleation (see section \ref{homn}).
\item If $\Delta_{\rm nuc} < \Delta_T^{\rm rms}$, the nucleation rate
      is inhomogeneous and we have to consider the scenario sketched in 
      fig.~\ref{fig8}. 
\end{enumerate}
A first attempt to analyse inhomogeneous nucleation has been given in \cite{IS}.
According to \cite{IS}, the nucleation distance $d_{\rm nuc}$ exceeds the 
scale $\lambda_{\rm smooth}$, if 
\begin{equation}
\lambda_{\rm smooth} < 2 {v_{\rm heat}\over 3 c_s^2} \Delta_T^{\rm rms}
R_{\rm H}.  
\end{equation}
If $\Delta_T^{\rm rms} > 5 \times 10^{-5}$, it is quite likely that this 
condition is met. In that case 
we can conclude that the typical inhomogeneity scale in the baryon 
distribution is inherited from the scale of density inhomogeneities in 
the radiation fluid at the end of the QCD transition. The effect in terms
of length scales is at least two orders of magnitude larger than the 
nucleation distance in homogeneous nucleation and is  
${\cal O}(1 \mbox{\ m})$, which is of interest for inhomogeneous BBN.
 
\section{Density fluctuations of the radiation fluid \label{cosmpert}}

\subsection{Amplification of fluctuations \label{radpert}}

Since the speed of sound vanishes during a first-order QCD transition 
(see section \ref{sound}), the restoring forces for compressional 
perturbations vanish and thus density inhomogeneities on scales below the 
Hubble scale are amplified \cite{SSW,SSW2,SW}. For small perturbations 
$\epsilon({\bf x},t)=:\epsilon_0(t)+ \delta \epsilon({\bf x},t)$, the 
equations of motion can be linearized in the perturbations.
Here we are interested 
in the density perturbations, the quantity of interest is the density 
contrast $\delta \equiv \delta \epsilon({\bf x},t)/\epsilon_0(t)$. 

The transfer functions, i.e.\ the change in the primordial spectrum,
for the radiation fluid and the cold dark matter (CDM) are calculated in
\cite{SSW,SSW2}. 

\begin{figure}[t]
\centerline{\includegraphics[width=0.65\textwidth]{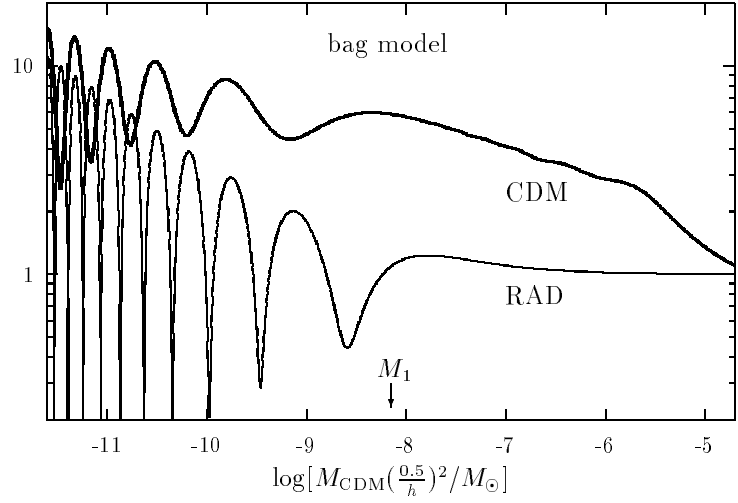}}
\caption{\label{fig9}
The modifications of the density contrast for kinetically decoupled CDM (e.g.
axions or primordial black holes), $A_{\rm cdm} \equiv | \delta_{\rm cdm}|
(T_\star /10)$, and of the radiation fluid amplitude, $A_{\rm rad} \equiv
(\delta_{\rm rad}^2 + 3 \hat{\psi}_{\rm rad}^2)^{1/2}$, due to the QCD
transition in the bag model (from \cite{SSW2}). Both quantities are normalized 
to the pure
Harrison--Zel'dovich radiation amplitude. On the horizontal axis the wave number
$k$ is represented by the CDM mass contained in a sphere of radius $\pi/k$.}
\end{figure}
For the bag model the transfer functions are shown in fig.~\ref{fig9}. Both
transfer functions show huge peaks on small scales. The different scales
$k_{\rm ph}=2\pi/\lambda$ are represented by the invariant CDM mass 
contained in a sphere of radius $\lambda/2$,
\begin{equation}
M_{\rm cdm}(\lambda)\equiv \frac{4 \pi}{3}\rho_{\rm cdm}
\left(\frac{\lambda}{2}\right)^3 \ ,
\end{equation}
assuming  for simplicity that $\Omega_{\rm cdm} \approx 1$ today. The 
largest scales in fig.~\ref{fig9} correspond to the
horizon at $T_{\rm f}=T_{\star}/10$. The CDM curve also shows the
logarithmic growth of subhorizon scales of CDM in a radiation-dominated
Universe. The CDM peaks lie on top of this logarithmic curve.

The peak structure starts at a scale $\approx 10^{-8} M_{\odot}$ in CDM
mass. This scale corresponds to the horizon scale at the QCD transition. The
radiation energy inside the horizon at $T_{\star}$ is $\sim 1 M_{\odot}$, but
it is redshifted as $M_{\rm rad}(a) \sim (a_{\rm eq}/a) M_{\rm cdm}$. 
Scales which are above the horizon at the QCD transition are not
affected. For scales below  $M_1^{\rm cdm} \approx 9 \times 10^{-9}
M_{\odot}$ the radiation peaks grow linearly in wave number.
This linear growth comes from the fact that the vanishing speed of sound during
the QCD transition implies a vanishing restoring force in the acoustic
oscillations on subhorizon scales. Therefore, the radiation fluid falls freely
during the transition, with a constant velocity given at the beginning of the
transition. The density contrast $\delta^{\rm rad}$ grows linearly in time
with a slope $k$. CDM is moving in an external potential provided by the
dominant radiation fluid, and is pushed by the strong increase in the
gravitational force during the transition. The highest peaks have
$k/k_1 \sim 10^4$, because on smaller scales the acoustic oscillations are
damped away by neutrino diffusion already before the QCD transition
(see section \ref{nudamp}).

\begin{figure}[t]
\centerline{\includegraphics[width=0.65\textwidth]{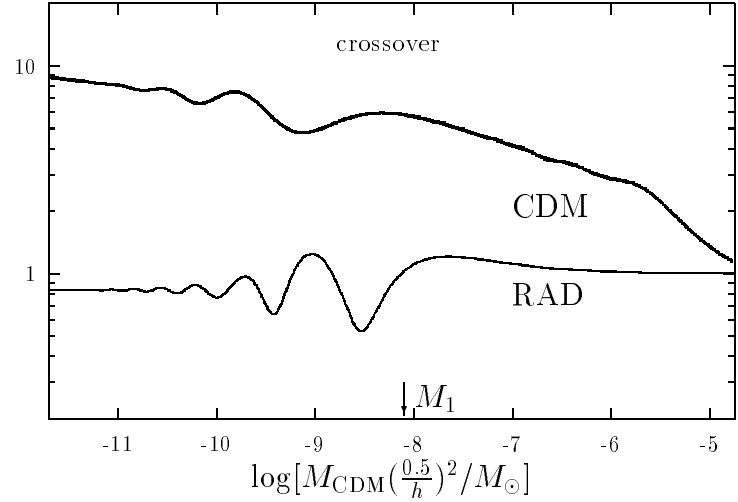}}
\caption{\label{fig10}
The same as fig.~\ref{fig9}, but for a QCD crossover 
(from \protect\cite{SSW2}).}
\end{figure}
The processed spectrum for a crossover, fig.~\ref{fig10}, shows a
behaviour similar to that for the bag model on superhorizon and horizon 
scales. The peak structure starts at $M_1$, but on subhorizon scales
there are no peaks. The level of the subhorizon transfer function for the
radiation fluid is reduced to $0.83$. This comes from the damping of
the acoustic oscillations during the time with $c_s^2 \ne 1/3$.

The time evolution of subhorizon modes, $\lambda \ll R_{\rm H}$,  
can be solved analytically during the transition. For the dynamics of
the radiation fluid (QCD, photons, leptons) CDM can be neglected, 
since $\epsilon_{\rm cdm}/\epsilon_{\rm rad} = a/a_{\rm eq} \approx
10^{-8}$. The transition time is short with respect to the Hubble time at the
transition, $(t_+ - t_-) < t_{\rm H}=H^{-1}$. For subhorizon modes we can
neglect gravity during the whole transition, as has been shown in \cite{SSW2}.
The damping terms in the continuity equation and Euler equation are
absent in the purely radiation dominated regime.
During the transition the damping terms can be neglected in view of
the huge amplification for a first-order phase transition. The continuity and
Euler equations read:
\begin{eqnarray}
\label{psihat}
  \delta' -k \hat{\psi}  = 0 \\\nonumber
  \hat{\psi}' + c_s^2 k\delta=0 \ ,
\end{eqnarray}
where $\hat{\psi}$ denotes the peculiar velocity up to a factor $\epsilon/
(\epsilon + p)$. The primes denote derivatives with respect to conformal time.
Written as a second-order differential equation for 
$\delta$, this is just an oscillator equation
\begin{equation}
\label{oscillator}
  \delta '' + \omega^2 \delta =0 \ ,
\end{equation}
with the time-dependent frequency $\omega(\eta) \equiv k c_s(\eta)$.

\begin{figure}
\centerline{\includegraphics[width=0.65\textwidth]{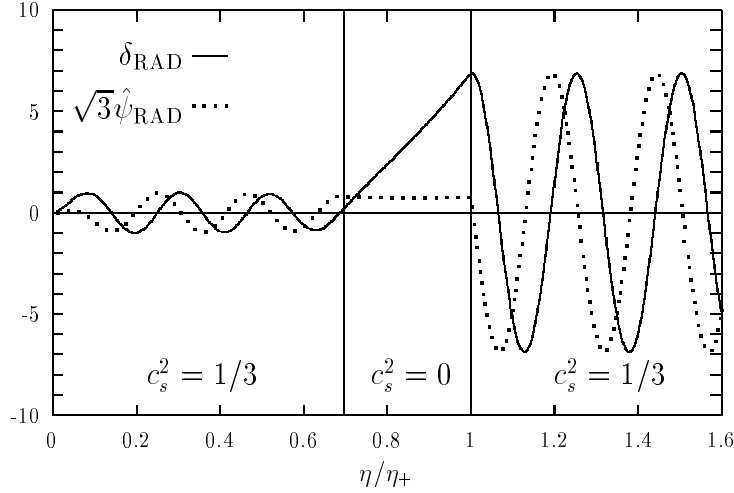}}
\caption{\label{fig11}
The time evolution of the density contrast $\delta_{\rm rad}$
and the peculiar velocity $\frac34 \hat{\psi}_{\rm rad}$
of the radiation fluid for the mode $k/k_1=7$ (from \protect\cite{SSW2}).
During the QCD transition in the bag model ---
marked by the 2 vertical lines --- the velocity stays
approximately constant and the density contrast grows linearly.
The amplitude is normalized to $1$ long before the transition.}
\end{figure}

Let us discuss the origin of large amplifications for small scales in the
bag model. Before and after the QCD transition, the radiation fluid
makes standing acoustic oscillations in each mode $k$, with speed of sound 
$c_s^2 =
1/3$ and with amplitudes $A_{\rm in}$ and $A_{\rm out}$ for the incoming and
outgoing solution respectively, see fig.~\ref{fig11}. The incoming solution for
the density contrast $\delta$ and the peculiar velocity $\sqrt{3}\hat{\psi}$
reads
\begin{eqnarray}
 \delta &=&-A_{\rm in} \cos{[\omega(\eta-\eta_-)+\varphi_-]} \\ \nonumber
 \sqrt{3}\hat{\psi}&=&A_{\rm in} \sin{[\omega(\eta-\eta_-)+\varphi_-]} \ .
\label{bagin}
\end{eqnarray}
This solution is valid until the beginning of the transition at
$\eta=\eta_{-}$, and $\varphi_-$ denotes the phase of the oscillation at
$\eta_-$. During the transition the speed of sound is zero. There are no
restoring forces from pressure gradients and the radiation fluid falls
freely. Since the duration of the transition is short with respect to the 
Hubble time $\Delta t\equiv (t_+ - t_-) < H^{-1}$, gravity is negligible 
during this free fall. The fluid is thus moving 
inertially in the sense of Newton, the velocity stays constant, and the density
contrast grows linearly in time:
\begin{eqnarray}
 \delta &=& \delta_-+k(\eta-\eta_-) \hat{\psi}_- \ ,
 \label{freefall} \\ \nonumber
 \sqrt{3}\hat{\psi}&=&\sqrt{3}\hat{\psi}_- \ ,
\end{eqnarray}
where $\sqrt{3}\hat{\psi}_-=A_{\rm in} \sin{(\varphi_-)}$ is the peculiar
velocity at $\eta_-$.

Since we have no jumps in pressure, the density contrast $\delta$ and the 
fluid velocity $\hat{\psi}$ stay continuous throughout the whole transition, 
in particular at the matching points of the different regimes. Gravity 
remains negligible during the entire transition.
At the end of the transition this solution has to be
joined to the pure radiation-dominated regime for $T\le T_{\star}$. Since the
amplitude of the density contrast grows linearly during the transition, the
final amplitude $A_{\rm out} = A_+$ is enhanced linearly in $k$,
modulated by the incoming phase
\begin{equation}
\label{k1}
  \left(\frac{A_{\rm out}}{ A_{\rm in}}\right)^2 \!\!\!=
\left({k \over k_1}\right)^2 \sin^2{(\varphi_-)} \ ,
\end{equation}
with $k_1\equiv \sqrt{3}/\Delta \eta$, $\Delta \eta \equiv \eta_+-\eta_-$.
The envelope of the linearly growing peak structure for subhorizon scales
starts at the scale $k_1$, which corresponds to a CDM mass of $M_1 = 9 \times
10^{-9} M_{\odot}$.

In the case of a crossover, the amplification occurs for scales
around the Hubble radius at the transition only. Subhorizon scales always
stay in the WKB-regime, and therefore the spectrum is flat for these scales.
However, the amplitude for subhorizon scales is damped during the
phase transition. The same damping occurs in the case of a first-order
phase transition. It has been neglected in the analytic discussion, since
it is a small correction. The subhorizon amplitudes are reduced to $83$\% 
of their initial value \cite{SSW2}. 

\subsection{Formation of black holes? \label{bh}}

It was suggested in the literature \cite{Crawford,Jedamzik,Jedamzik2} that
the QCD transition could lead to the formation of $1 M_{\odot}$ black holes,
which could account for today's dark matter. Jedamzik \cite{Jedamzik}
proposed to identify such black holes with the MACHOs observed by
microlensing \cite{microlensing}. He pointed out that the formation of black
holes should be particularly efficient during the QCD epoch thanks to the
significant decrease in the effective speed of sound.

In order to form a black hole in a radiation-dominated
Universe, the density contrast inside the Hubble radius should be in the
range $1/3 <\delta_{\rm H}<1$ \cite{BHF}. For an  observable amount of $1
M_{\odot}$ black holes today, i.e. $\Omega_{\rm BH}^{(0)}={\cal O} (1)$, the
fraction of energy density converted to black holes at the QCD transition must
be ${\cal O} (a_{\rm QCD}/a_{\rm eq}) \approx 10^{-8}$. For a Gaussian
distribution this requires $\delta_{\rm rms} \approx 0.06$ (without including
any enhancement from the QCD transition) \cite{Bullock}. The QCD transition
gives an enhancement factor (at the horizon scale) of $2$ for the bag model and
of $1.5$ for lattice QCD in the linear perturbation treatment,
figs.~\ref{fig9} and \ref{fig10}. This indicates a corresponding reduction in
the required pre-existing perturbation spectrum at the solar mass scale.
Cardall and Fuller used a qualitative argument of Carr and Hawking
\cite{CarrHawking} and the bag model; they also obtained a reduction
of a factor of $2$ in the required pre-existing perturbation spectrum 
\cite{Cardall}. These QCD
factors of $1.5$ or $2$ are so modest that a pre-existing Harrison--Zel'dovich
spectrum with COBE normalization is very far from giving a cosmologically
relevant amount of black holes \cite{SW}. One would have to put in a
fine-tuned tilt  $(n-1) \approx 0.4$ to get the desired amount of black
holes. However, this tilted spectrum would overproduce primordial black holes
on scales that are only a factor of $50$ below the Hubble radius at the QCD
transition. Therefore a break in the pre-existing spectrum below the QCD scale
would be required, a second fine-tuning.

We conclude that the QCD transition does not produce black holes, although 
it enhances the probability to form some. The pre-existing spectrum needs 
to be fine-tuned around the QCD scale if one would like to produce a 
detectable amount of black holes. Thus the major effect would not be due to 
the QCD transition, but due to a feature in the primordial spectrum.

\subsection{Collisional damping at neutrino decoupling \label{nudamp}}

The acoustic oscillations in the radiation fluid get damped by neutrino
diffusion at the time of neutrino decoupling. This damping is analogous to
Silk damping at photon decoupling. The muon and tau neutrinos, which are
coupled to the radiation fluid via neutral current interactions only, 
decouple at
$T^{\rm dec}_{\nu_\mu \; \nu_\tau} \sim 2.2$ MeV from the Hubble scale $R_{\rm
H}$, which follows from Ref.~\cite{Heckler}. The electron neutrinos interact
by charged and neutral currents and decouple slightly later, $T^{\rm
dec}_{\nu_e} \sim 1.4$ MeV. By the time of neutrino decoupling at the Hubble
scale all inhomogeneities in the radiation fluid on scales below 
$\approx 10^{-6} M_{\odot}$ in CDM mass are wiped out by neutrino diffusion 
(cf.~the QCD horizon scale is $10^{-8} M_\odot$ in CDM mass), as shown below,
eq.~(\ref{nudamping}). Therefore the QCD peaks cannot affect big bang
nucleosynthesis (BBN).

It is important to distinguish the total decoupling of neutrinos,
i.e. a neutrino decoupling at the Hubble scale, when the neutrino
scattering rate $\Gamma$ is smaller than the Hubble rate, $\Gamma_{\nu}/H<1$,
from the neutrino decoupling with respect to a
certain mode given by  $\Gamma_{\nu}/\omega_{\rm ph}<1$, when the typical
neutrino scatters less than once during an acoustic oscillation time of one
particular mode. The mode-dependent decoupling temperature $T_{\nu}^{\rm
dec}(k)$ is related to the total decoupling temperature by
\begin{equation}
{T^{\rm dec}_{\nu_i}(k) \over T^{\rm dec}_{\nu_i}(H)} \approx \left({ c_s
k_{\rm ph}\over H}\right)^{1/4}_{T=T^{\rm dec}_{\nu_i}(H)} \ .
\end{equation}
This is because the neutrino interaction rates $\Gamma_{\nu}$ are
proportional to $T^5$ and $k_{\rm ph} \propto 1/a \propto T$, hence
$\Gamma_{\nu}/\omega_{\rm ph} \propto T^4$.

To compute the damping of acoustic oscillations in the radiation fluid by
neutrino diffusion, we follow Weinberg \cite{Weinberg}. For a radiation fluid,
shear viscosity is dominant,  bulk viscosity vanishes  and heat conduction 
is negligible. The shear viscosity is given by
\begin{equation}
 \eta_{\rm visc}={4 \over 15} \sum_i \epsilon_{\nu_i} \tau_{\nu_i} \ ;
\end{equation}
$\epsilon_{\nu_i}$ denotes the energy density of a neutrino species,
$\tau_{\nu_i}$ is the typical collision time. In the subhorizon limit the
Navier--Stokes equation and the continuity equation give
\begin{equation}
  \delta'' + {k_{\rm ph} \over \epsilon_{\rm tot}} \eta_{\rm visc} k \delta' +
  \omega^2 \delta = 0,
\end{equation}
a damped oscillator. The damping factor for the mode $k$ at a given conformal
time $\eta$ is
\begin{equation}
\label{dampingfactor}
D(k,\eta)= \exp{\left[-{1 \over 2}\int_0^{\eta_{\rm max}}
 (k_{\rm ph} / \epsilon_{\rm tot}) \eta_{\rm visc} k d \eta'\right]}.
\end{equation}
Here the upper limit of the integral is the conformal time $\eta_{\rm max} =
{\rm min}[\eta,$ $\eta_{\rm dec}(k)]$, because
collisional damping of the $\gamma$--$\ell^{\pm}$--hadron fluid by neutrinos
ceases at the decoupling of the mode $k$ considered. The damping per oscillation
is largest for $\omega \tau_{\nu} \equiv \omega/\Gamma_{\nu} \approx 1$,
i.e. immediately before neutrino decoupling for a given  mode. But note
that subhorizon modes get strongly damped long before the mode decouples from
neutrinos, because a weak damping per oscillation is compensated by many
oscillations per Hubble time.

For a first application, we ask what modes are already damped by the time of
the QCD transition. At the QCD transition, $T=T_{\star}$, the interaction rates
for electron and muon neutrinos (resp. antineutrinos) with the leptons are the
same, $\Gamma_{\nu_e}=\Gamma_{\bar{\nu}_e}=\Gamma_{\nu_\mu}=
\Gamma_{\bar{\nu}_\mu} =3.1 G_{\rm F}^2 T^5$ \cite{Heckler},  since electrons
and muons are still  relativistic. The $\tau$-neutrinos interact only via
neutral currents with the leptons and have a lower interaction rate,
$\Gamma_{\nu_\tau}= \Gamma_{\bar{\nu}_\tau} = 0.6 G_{\rm F}^2 T^5$.
Evaluating the damping integral, eq.~(\ref{dampingfactor}), at $T=T_\star$ we
find that the damping factor $D(k,T_\star)$ is $<1/e$ for $(k_{\rm
ph}/H)_{T_{\star}} > 10^4$, which corresponds to $M_{\rm cdm} <10^{-20}
M_{\odot}$, i.e. acoustic oscillations on these small scales are wiped out
before the QCD transition. Therefore no peaks in the radiation or in the CDM
transfer function can develop below this scale. This small-scale cut-off is
independent of the bubble separation scale.

Next we consider $T<T_{\nu_e}^{\rm dec}(H) \approx 1.4$ MeV and evaluate the
final damping factor $D(k)$.
At the time of neutrino decoupling we take a purely radiation-dominated
Universe, consisting of only $\gamma$, $e^{\pm}$ and $\nu$ to evaluate the
damping. Muons have disappeared since $m_{\mu} \gg T$ and since $\tau_{\mu}
\ll t_{\rm H}$. The interaction rate for electrons is still given by charged
and neutral currents, $\Gamma_{\nu_e}=\Gamma_{\bar{\nu}_e}=1.3 G_{\rm F}^2
T^5$, whereas muon and tau neutrinos have the same lower (neutral current)
rate, $\Gamma_{\nu_\mu}=\Gamma_{\bar{\nu}_\mu}=\Gamma_{\nu_\tau}=
\Gamma_{\bar{\nu}_\tau}=0.3 G_{\rm F}^2 T^5$ \cite{Heckler}. The final damping
of a certain scale (expressed in invariant CDM mass) from
eq.~(\ref{dampingfactor}) is
\begin{equation}
\label{nudamping}
D = \exp{\left[-\left({M_{\nu-{\rm dmp}} \over M}\right)^{1/4}\right]},
\end{equation}
with the neutrino damping scale $M_{\nu-{\rm dmp}} \approx 1.9\times 10^{-6}
M_{\odot}$ in CDM mass. This is $3 \times 10^{-5} M_{\rm H}^{\rm cdm}$ at $T =
1.4$~MeV and corresponds to length scales $\lambda = 1/30 R_{\rm H}$.

We conclude that any small-scale density (temperature) fluctuations of the 
radiation fluid are damped away by the time of BBN. The large peaks
in the radiation fluid that could be generated by the QCD transition 
do not give rise to inhomogeneous BBN. 

\section{Isothermal baryon fluctuations \label{baryonfluc}}

The nucleation of bubbles in a first-order QCD transition may strongly 
affect the distribution of baryons in the Universe \cite{Witten,Applegate}. 
Isothermal baryon number fluctuations are created at the end of the QCD 
phase transition. These fluctuations are smoothened 
by baryon diffusion after the QCD transition. As the  
temperature of the Universe drops below $1$ MeV, neutrons and protons 
decouple chemically (freeze out). The mean free path of the neutrons 
exceeds that of the protons and protons and neutrons are thus segregated. 
In other words, regions of high baryon density become proton-enriched and 
regions of low baryon density become neutron-enriched. 
This affects the local nucleosynthesis yields, giving rise to inhomogeneous 
big bang nucleosynthesis (IBBN) \cite{IBBN} (in contrast to homogeneous 
BBN \cite{BBN}), which is sketched in section \ref{ibbn}. 

\subsection{Baryon inhomogeneities \label{binhom}}

Isothermal fluctuations of baryon number are an inevitable consequence of any
first-order QCD transition in the early Universe. After the tiny period of 
supercooling, bubble nucleation, and reheating, the HG bubbles grow 
adiabatically and merge (percolate). For tiny supercooling the percolation 
happens very close to equilibrium, because 
this process is extremely slow with respect to the QCD time scale. Just before 
the transition is completed, the Universe is filled with a phase of HG in 
which shrinking drops of QGP are embedded. Usually these drops are 
assumed to be spheres, which is the case if the surface tension is reasonably 
large. In the homogeneous nucleation scenario, the typical distance 
between QGP drops is given by the typical bubble nucleation distance 
$d_{\rm nuc}$. The bulk properties of growing HG bubbles 
and shrinking QGP drops (except for the very last moments before the drops
evaporate)
are governed by the hydrodynamic equations for the radiation fluid only,
i.e.~the baryons do not influence the bulk properties. The hydrodynamics of 
growing HG bubbles and decaying QGP drops has been studied in detail in 
Refs.~\cite{Kurki-Suonio,Kajantie,Kurki-Suonio96}. Without baryons, no relics
would remain from the bubble scale today, so let me concentrate on the 
role of baryons during a first-order QCD transition.

Let me assume for a while that chemical and thermal equilibrium are maintained
from the formation of the QGP drops until the completion of the phase 
transition (evaporation of the last QGP drops). In chemical equilibrium the 
baryon number prefers to reside in the QGP, since the deconfined quarks are 
almost massless. The ratio $\kappa$ of the baryon number density in the two
phases is a constant for chemical equilibrium, given by
\begin{eqnarray}
\label{kappa}
\kappa &\equiv  & {n_{\rm B}^{\rm QGP}\over n_{\rm B}^{\rm HG}} \\
       &\approx &
        {N_f\over 9} \left({\pi\over 2}\right)^{\frac32}  
        \left(\frac{m_{\rm p}}{T_\star}\right)^{-\frac32} 
        \exp\left(\frac{m_{\rm p}}{T_\star}\right) \\
       &\approx & 16 \qquad \mbox{for\ } T_\star = 160 \mbox{\ MeV\ }, 
\nonumber
\end{eqnarray}
evaluated for the bag model with $N_f = 3$. The mean baryon number density 
is given by
\begin{equation}
\label{meannB}
n_{\rm B} = (1 - f) n_{\rm B}^{\rm QGP} + f n_{\rm B}^{\rm HG} \ ,
\end{equation}
where $f$ is the volume fraction filled by the HG. It rises from 
$0$ to $1$ during the transition. After reheating, $f \sim 10^{-3}$. At 
percolation $f \sim 0.7$ (for spherical bubbles with a single radius) 
\cite{perlocation}. The baryon number $N_{\rm B} = n_{\rm B} V$ is 
conserved in the comoving volume $V$. 
{}From eqs.~(\ref{kappa}) and
(\ref{meannB}) one finds that both $n_{\rm B}^{\rm QGP}/n_{\rm B}$ and 
$n_{\rm B}^{\rm HG}/n_{\rm B}$ grow during the transition:
\begin{eqnarray}
\label{nBQGP}
n_{\rm B}^{\rm QGP} &=& {n_{\rm B}\over f/\kappa + (1-f)} \\ 
\label{nBHG}
n_{\rm B}^{\rm HG}  &=& {n_{\rm B}\over f + (1-f)\kappa} \ .
\end{eqnarray}
The maximum values are reached for $f \approx 1$, where $n_{\rm B}^{\rm QGP}/
n_{\rm B} \approx \kappa = {\cal O}(10)$ and $n_{\rm B}^{\rm HG}/n_{\rm B} 
\approx 1$. Of course, if 
equilibrium were maintained throughout the transition, no baryon inhomogeneity 
would arise from the preference of baryons to reside in the QGP. If baryons
drop out of equilibrium before the transition is completed, baryon 
inhomogeneities are necessarily produced. 

There are three mechanisms that drive the baryons out of chemical equilibrium:
\begin{itemize}
\item The baryon flux through the surface of the QGP drop might be suppressed.
This was elaborated in Refs.\ \cite{Applegate,Fuller,Sumiyoshi}. Any 
suppression of the baryon flux
results in a deviation from chemical equilibrium. Since the net flow of baryons 
is directed from inside the QGP drops to the HG, flux suppression increases
the value of $\kappa$ and baryon number fluctuations are generated.

However, we do not know the physics of the phase interface so far. It is thus
unclear how large this flux suppression might be. The simplest model is 
based on phase-space arguments \cite{Fuller}. A more sophisticated model 
attaches chromoelectric flux tubes to the surface at the points where a quark 
penetrates the interface. These flux tubes then break up and produce mainly 
pions, and rarely nucleons or heavier baryons \cite{Sumiyoshi}. Both models 
give a large suppression of baryon number flux. However, in both models it 
is assumed that no bound states or cluster states exist in the QGP near the 
critical temperature. If, for instance, a significant fraction of quarks in the QGP 
would form diquarks near the critical temperature, the arguments for flux 
suppression would break down. 

Our ignorance about the physics of the phase interface has been encoded 
in the so-called filter factor~$F$ \cite{Fuller}, which is any number 
between $0$ and $1$
and gives the probability that a baryon penetrates the interface. The flux 
suppression is strongest and most simple to analyse when $F=0$, which means 
that the baryon number resides in the QGP phase. In such a situation 
the only relevant quantity is the typical nucleation distance. Any QGP drop 
contains a baryon number of about $N_{\rm B} = n_{\rm B} d_{\rm nuc}^3$, 
which gives 
$\sim 10^{30}\, (10^{36})$ baryons per QGP drop for $d_{\rm nuc} \sim 10^{-6} 
\, (10^{-4}) R_{\rm H}$ after the annihilation of all antibaryons in the QGP 
drop. The baryon density in the QGP drops reaches nuclear matter density 
($\sim 200$ MeV/fm$^3$) 
at a drop radius of $10\ \mu\mbox{m}\, (1 \mbox{\ mm})$. 
This demonstrates that the produced inhomogeneities may be enormous. On the
other hand it is clear that the filter factor cannot vanish exactly, 
because all 
baryons would be trapped in the QGP drops (quark nuggets) by the end of the 
QCD transition and no nucleosynthesis could have taken place. 

\item Even if the baryon flux through the phase interface is large enough to
maintain chemical equilibrium locally, we must know the diffusion 
length of baryons during the evaporation of the QGP. Only if baryons 
diffuse over scales $d_{\rm nuc}$ can chemical equilibrium be 
maintained globally. According to eq.~(\ref{nBHG}), $n_{\rm B}^{\rm HG} \approx
n_{\rm B}/\kappa \ll n_{\rm B}$ at the early stages of the transition, and 
thus the regions where HG bubbles nucleate are underdense in baryon number. 

The typical collision time for baryons is $\tau_{\rm b} \sim 1/T_\star$ 
(collisions with $\pi$s dominate at $T_\star$). The duration of the transition 
is $\sim 0.1 t_{\rm H}$. Thus, baryons make $N \sim 10^{19}$ collisions during 
the phase transition. Their typical velocity is $v_{\rm b} \sim 
(T_\star/m_{\rm p})^{1/2}$ and thus they diffuse a distance
of $\lambda_{\rm b-diff} \sim \sqrt{N} v_{\rm b} \tau_{\rm b} \sim 
10^{-11} R_{\rm H}$, which is much smaller than $d_{\rm nuc}$. Thus baryon 
diffusion alone is not able to keep the mixed phase in global chemical 
equilibrium. 

The consequences of baryon diffusion have been studied by Kurki-Suonio
\cite{Kurki-Suonio88} and have been incorporated into numerical simulations
\cite{Rezzolla}. Since $R_{\rm drop} \gg \lambda_{\rm b-diff}$ initially,
chemical equilibrium cannot be established globally (not even within a single 
phase). At the QGP side of the interface, the baryon number is piled up in a 
thin
surface layer of dimension $r_{\rm s}\sim \lambda_{\rm b-diff}^2/R_{\rm drop}$ 
\cite{Kurki-Suonio88}. Since locally $n_{\rm B}^{\rm QGP} = 
\kappa n_{\rm B}^{\rm HG}$, more baryons leak into the HG than without 
diffusion. The baryon density grows when the QGP drops shrink, and thus 
the baryon density in the HG close to the QGP drop grows 
as well. Since the diffusion length is small the baryons that penetrate 
the interface cannot be redistributed in the whole volume. In this way 
core-concentrated clouds of baryons are formed. 

\item The third mechanism is the extraction of entropy once the mean free path 
of any relativistic species exceeds the size of the QGP drops. This 
happens first for the neutrinos, $\lambda_{\nu-{\rm mfp}} \sim 
10^{-6} R_{\rm H}$, later for the charged leptons. A radiative flux is 
established, which carries 
away entropy but not baryon number. In contrast the hydrodynamic flux 
carries away entropy as well as baryon number. Thus baryons are compressed in 
the QGP drop (it is the ratio $n_{\rm B}/s$ that is important). This 
mechanism has been found by Applegate and Hogan \cite{Applegate} and was 
further studied in \cite{Kurki-Suonio88,Fuller,Rezzolla}. It turns out that
entropy extraction is most efficient for large initial drop size 
($R_{\rm drop} \gg \lambda_{\nu-{\rm mfp}}$).
\end{itemize}
All three mechanisms lead to the formation of baryon number fluctuations by the 
end of a first-order QCD transition. Because the baryons are tightly coupled
to the radiation fluid on scales of ${\cal O}(d_{\rm nuc})$, the baryon fluid
has the same local temperature as the radiation fluid and thus isothermal
fluctuations are generated (in contrast to isentropic fluctuations). Let me
point out that the second and third mechanisms are independent from the 
details of the QCD phase transition, especially from its order. It might 
therefore be possible that an inhomogeneous (temperature fluctuations 
from cosmological inflation) crossover could introduce baryon inhomogeneities 
as well. The typical length scale would be $\lambda_{\rm smooth}$, as in 
the case of inhomogeneous nucleation. To my knowledge, this issue has not 
been investigated in any detail yet.

Let me estimate (for a first order transition) the fluctuation amplitude 
and the fraction of baryons affected:
the most extreme case takes place for the observationally excluded situation 
of a vanishing filter factor ($F=0$). For small values of $F$ we still find 
extremely high baryon densities
and a large fraction of affected baryons. Given the large uncertainties on the
value of $F$, it is most interesting to ask what happens if $F=1$. Since the 
physics of baryon diffusion and entropy extraction are well understood, this
case gives a lower limit on the fraction of baryons that is concentrated in 
clouds. I will focus on large nucleation distances, I take $d_{\rm nuc} 
\sim 10^{-4} R_{\rm H}$, because, as will be shown below, 
$d_{\rm nuc} \geq 1$ m is necessary in order to affect BBN.

The first substantial departure from chemical equilibrium happens when 
$R_{\rm drop} \sim \lambda_{\nu-{\rm mfp}} \sim 10^{-6} R_{\rm H}$. 
For $d_{\rm nuc} \sim 10^{-4} R_{\rm H}$, the volume fraction in QGP 
drops at $R_{\rm drop} \sim \lambda_{\nu-{\rm mfp}}$ is $f_{\rm QGP} 
\sim 10^{-6}$. Since at that time 
$\lambda_{\rm b-diff} \ll R_{\rm drop}$ the fraction of baryons that is piled 
up in a surface layer on the QGP side of the interface is negligible.
This means that the fraction of baryons that is inside the QGP drops at
neutrino decoupling is $\sim \kappa f_{\rm QGP}$, according to 
eq.~(\ref{nBQGP}), and therefore at most $\sim 10^{-5} N_{\rm B}$ 
baryons are in 
core-concentrated clouds by the end of the QCD transition. 

It was shown in Refs.~\cite{Applegate,Kurki-Suonio88} that for
$R_{\rm drop} \gg \lambda_{\rm b-diff}$ the baryon 
number distribution of such a cloud is given by
\begin{equation}
\label{nr}
n_{\rm B}(r) \approx \bar{n}_{\rm B}\left[
\left({R_{\rm cloud}\over r}\right)^\alpha - 1\right] \ ,
\end{equation}
where $R_{\rm cloud} \sim \lambda_{\nu-{\rm mfp}}$ and $\alpha$ is the ratio
of the neutrino radiation flux and the hydrodynamic flux, i.e.
\begin{equation}
\alpha \equiv {F_{\nu}\over F_{\rm hydro}} \propto 
\sqrt{\Delta T\over T_\star} \ll 1 \ .
\end{equation}
This ratio depends on the temperature difference between the QGP drop
and the surrounding HG. The temperature gradient is necessary to establish a 
hydrodynamic flux. This flux originates from a surface 
layer, whereas the neutrino flux originates from the volume of the drop. 
When the 
QGP drop shrinks, the ratio $\alpha$ grows because $\Delta T$ has to
grow to keep both phases close to $T_\star$. Within the bag model
$\alpha \sim 10^{-4}$ at most. The parameters from quenched 
lattice QCD give $\alpha \sim 10^{-6}$. The resulting baryon 
distribution (\ref{nr}) is very broad and high baryon densities are 
reached at the centre of the clouds only. The fraction of baryons in the 
clouds has been estimated to be $\alpha/3$ \cite{Kurki-Suonio88}, which is 
consistent with the estimate $\sim \kappa f_{\rm QGP}$. In general, 
increasing the nucleation distance reduces the fraction of baryons in high
density regions. 

It was shown numerically in Ref.~\cite{Rezzolla} that enormous baryon 
densities are reached in the inner $10^4$ fm of the drop. However, it is hard
to have a cosmologically significant fraction of baryons residing in these
inner regions if there is no baryon flux suppression. Large 
separations of baryon clouds and high baryon fraction within these clouds
are mutually exclusive, except for significant baryon flux suppression.

\subsection{Baryon diffusion \label{bdiff}}

The isothermal baryon number fluctuations formed at the end of a first-order 
QCD transition are smooth\-en\-ed by two mechanisms, baryon diffusion 
\cite{Applegate87,Banerjee,Mathews} and neutrino inflation
\cite{Applegate,Heckler,Jedamzik94a}. The latter is important 
for baryon clouds with high $\eta \equiv n_{\rm B}/n_{\gamma} > 10^{-4}$. 
The pressure of these baryon clouds is in equilibrium with the regions of lower
baryon density. Thus the temperature of the baryons could be below the mean 
temperature, because the baryons give an additional contribution to the 
pressure besides the radiation quanta. If the size of the cloud is smaller than 
the mean free path of neutrinos, heat is transported into the baryon clouds.
This leads to an increase in pressure inside the cloud and thus to an expansion 
(inflation) of the cloud. Neutrino inflation provides an upper limit
$\eta \leq 10^{-4}$ for clouds with $R_{\rm cloud}\leq 1$ m at the time of 
the QCD transition. In contrast to neutrino inflation, baryon diffusion is an 
efficient smoothing mechanism for any value of $\eta$.

After the annihilation/decay of pions the baryon mean free path increases
(below $T \sim 40$ MeV). At temperatures above the decoupling of the weak
interaction ($\sim 1$ MeV) protons and neutrons are in chemical equilibrium 
and thus nucleons frequently change isospin. A nucleon spends a fraction 
$X_{\rm n}(T) \equiv n_{\rm n}/n_{\rm B} = 1/[1 + \exp(Q/T)]$ of its time 
as a neutron, where $Q = 1.29$ MeV is the neutron--proton mass difference. 
The mean free path of neutrons is 
much larger than the mean free path of protons. Thus at $T > 1$ MeV the 
diffusion length of nucleons is given by $\lambda_{\rm N-diff}(t) \sim 
\sqrt{\lambda_{\rm n-mfp} v_{\rm n} X_{\rm n} t}$, where $\lambda_{\rm n-mfp}$
is the neutron mean free path and $v_{\rm n}\sim (T/m_{\rm N})^{1/2}$ is the 
thermal velocity of neutrons. The neutron mean free path is determined from 
neutron--electron scattering (the neutron has an electric dipole moment) and 
from neutron--proton scattering. At $1$ MeV the nucleon diffusion length is 
$\approx 200$ m \cite{Applegate87,Mathews}. This corresponds to a comoving 
distance at the QCD transition of $\sim 1$ m or $\sim 10^{-4} R_{\rm H}$. 

After weak decoupling (at $\sim 1$ MeV) neutrons diffuse much larger 
distances than protons. The proton diffusion is determined by electron--proton
scattering, which gives a diffusion distance much smaller than that of the 
neutron. The proton diffusion at the time nucleosynthesis starts 
(at $\sim 0.1$ MeV) is completely negligible compared with the nucleon 
diffusion before weak decoupling. Thus $\lambda_{\rm p-diff} \sim 3$ km at 
the start of BBN and 
\begin{equation}
\lambda_{\rm p-diff} \sim 10^{-4} R_{\rm H} \mbox{\ comoving\ at\ } T_\star.
\end{equation}
The neutron diffusion length is $\lambda_{\rm n-diff} \sim 200$ km at
the start of BBN or
\begin{equation}
\lambda_{\rm n-diff} \sim 10^{-2} R_{\rm H} \mbox{\ comoving\ at\ } T_\star.
\end{equation}
The typical scale for baryon inhomogeneities are given by the typical 
bubble nucleation distance, $d_{\rm nuc}$.
Thus all baryon inhomogeneities on scales $d_{\rm nuc}< 
\lambda_{\rm p-diff}$ are washed out by baryon diffusion by the time 
BBN starts. Most interesting are baryon inhomogeneities on scales 
$\lambda_{\rm p-diff} < d_{\rm nuc} < \lambda_{\rm n-diff}$. The situation 
$d_{\rm nuc}>\lambda_{\rm n-diff}$ seems to be highly disfavoured, according to
the discussion in section \ref{bubbles}. Note that $R_{\rm H} \approx 
6 \times 10^7$ km at $T = 0.1$ MeV, thus $\lambda_{\rm p-diff} \sim 10^{-7}
R_{\rm H}$ at the start of BBN. 

The different diffusion lengths lead to the segregation of protons and 
neutrons. Neutrons escape from the high baryon density clouds,
while the protons remain in the cloud. Thus the surrounding cloud
becomes neutron-rich, whereas the baryon cloud itself becomes proton-rich.
This provides inhomogeneous initial conditions for nucleosynthesis. 

There is no damping mechanism for baryon inhomogeneities on scales 
larger $\lambda_{\rm n-diff}$. However, it is hard to imagine any mechanism 
that could have generated such fluctuations. As shown above, a first-order 
QCD transition would give rise to inhomogeneity scales just 
above $\lambda_{\rm p-diff}$. Especially, if we assume that 
cosmological inflation took place, baryogenesis must have happend after 
inflation and thus any baryon inhomogeneities must have been produced 
during the radiation-dominated epoch. Besides the cosmological QCD transition,
there is no other event before BBN that possibly could have generated 
isothermal baryon fluctuations at large enough scales.
If one assumes that the QCD transition does not give rise
to any relics, the baryon-to-entropy ratio is constant 
not only in time, but also in space.   
 
\subsection{Inhomogeneous nucleosynthesis \label{ibbn}}

The baryon segregation, which leads to an inhomogeneous neutron-to-proton 
ratio, and inhomogeneities in the entropy per baryon both lead to 
inhomogeneous nucleosynthesis \cite{IBBN,Jedamzik94b}. Let me first discuss 
what happens to the element abundances in a homogeneous patch when we 
change $\eta$ and $X_{\rm n}$. 
\begin{itemize}
\item[$\eta$]In homogeneous BBN, $\eta$ is the single free parameter of the 
theory. Based on measured deuterium abundances, we obtain 
$\eta = (5.6 \pm 0.5) \times 10^{-10}$ or $\omega_{\rm b} = 0.0205\pm 0.0018$ 
\cite{bbn}, consistent with the value obtained from CMB experiments 
\cite{cmb,Spergel}. The key role of 
$\eta$ is to determine the time when the nucleosynthesis starts. 
The first step in nucleosynthesis is the generation of deuterium
(p + n $\to$ d + $\gamma$). Since deuterium has a small binding energy
($\approx 2.2$ MeV) it is very easy to photo-dissociate it. The condition that
deuterium does not see photons with energy above $2.2$ MeV is $\eta^{-1} 
\exp(2.2 \mbox{\ MeV}/T) < 1$, which gives the temperature at the beginning
of nucleosynthesis. The higher $\eta$, the earlier BBN starts, which means 
that baryons can be burned more efficiently. For higher $\eta$,
more ${}^4$He and less D are produced. Typical mass fractions\footnote{Note, 
usually for D, ${}^3$He and ${}^7$Li the number fractions are given. Here 
I quote mass fractions for all elements.} 
in homogeneous BBN are $0.25, 5 \times 10^{-5}, 3 \times 10^{-5}, 
3 \times 10^{-9}$ for
${}^4$He, D, ${}^3$He, and ${}^7$Li, respectively. For very high baryon 
densities, i.e.~$\eta \sim 10^{-4}$, heavy elements such as ${}^{12}$C, 
${}^{13}$C, and ${}^{14}$N could be generated with mass abundances of 
${\cal O}(10^{-9})$ \cite{JedamzikHE}. 

\item[$X_{\rm n}$]The neutron--baryon fraction determines the 
mass abundance of ${}^4$He, given by $Y_{\rm p} \equiv 
4 n_{{}^4\rm He}/n_{\rm B}\approx 2 X_{\rm n}$. Here I used the approximation
that all neutrons eventually end up in ${}^4$He.
For homogeneous BBN $X_{\rm n} \approx 1/8$ and thus $Y_{\rm p} \approx 0.25$ 
from this simple argument. An inhomogeneous distribution 
of $X_{\rm n}$ could change the local ${}^4$He abundance. Of course, the other
abundances change as well.
\end{itemize}
In order to produce a global change in the elements abundances, non-linear 
fluctuations in $\eta$ and/or $X_{\rm n}$ are needed. Linear fluctuations 
would yield the same average abundances as homogeneous BBN. 
In addition the IBBN yields depend on the geometry of the baryon fluctuations.

Originally, the interest in IBBN arose because it was thought that 
$\Omega_{\rm B}=1$ was possible \cite{Applegate87}. It was shown that
even with IBBN a baryon-dominated, flat Universe is impossible, as it means 
a severe overproduction of ${}^7$Li \cite{Alcock87,Fuller,Jedamzik94b}.
Nevertheless there are interesting, probably in the future observable,
consequences from IBBN: the precision measurements of the primordial 
abundances will need to take effects from IBBN into account \cite{Kainulainen}
and, more speculative, higher mass elements (C, N, O) might have been 
generated during IBBN at an amount that could be observable \cite{JedamzikHE}.

The following conditions are necessary in order to have significant 
effects from IBBN: 1.~$d_{\rm nuc} > \lambda_{\rm p-diff}$, 
2.~$\eta_{\rm high}/\eta_{\rm low}$ should be large, and 3.~the fraction of 
baryons in the high density region has to be well above the ${}^7$Li 
abundance. Heavy elements could be produced, leaving the ${}^4$He abundance 
unchanged, if this 
fraction is small and $\eta_{\rm high}$ is at its upper limit $\sim 10^{-4}$. 
On the other hand, significant corrections to the light element yields are
obtained if the fraction of baryons in high density regions is large. 
In a recent work, Suh and Mathews \cite{Mathews} 
found that two fluctuation scales are 
consistent with the observed abundances: $< 1$ m (homogeneous BBN) and 
$10$--$40$ m, measured at $T_\star$. It seems that the nucleation distances
obtained from heterogeneous and inhomogeneous nucleation are consistent 
with these constraints on IBBN. 

\section{Cold dark matter\label{CDMQCD}}

The intention of this section is to highlight the differences between
various CDM candidates. By definition, CDM has negligible pressure and 
speed of sound when the formation of structure starts (at $t \sim t_{\rm eq}$). 
Before matter-radiation equality, inhomogeneities in the density of CDM grow 
logarithmically, whereas thereafter their growth is linear in the  
scale factor. This is the same for all CDM candidates, and thus they cannot 
be distinguished at mass scales above 
$M_{\rm eq} = 1.32 \times 10^{17} (0.15/\omega_{\rm m})^2 M_\odot$, 
corresponding to a comoving distance $R_{\rm eq} = 
91 (0.15/\omega_{\rm m})$~Mpc today. However, the astrophysics of CDM 
depends on the nature of CDM at scales well below $R_{\rm eq}$. 
Besides behaving like a non-relativistic 
fluid at large scales, a good CDM candidate should be inert to 
electromagnetic and strong interactions. Its interactions should not be  
stronger than the weak interaction.

Here I pick three examples: the neutralino, the axion and primordial black 
holes. The neutralino is an example of a WIMP. The general features of a WIMP 
are that it is massive ($m \gg T_{\rm eq}$) and interacts weakly. The axion
interacts weaker than the weak interactions of the standard model and it is 
very light. Nevertheless, it is a CDM candidate, because it oscillates 
coherently and axion strings decay into non-thermal axions. 
Primordial black holes are very 
heavy on the particle physics scale and their sole interaction is gravity.
Before discussing these candidates in more detail, let me mention some of
the other candidates that have been proposed (strangelets, quark nuggets 
and QCD-balls are discussed in section \ref{QCDoverview}). 

Another CDM candidate is a heavy neutrino (4th generation) 
with $m_Z /2 < m_\nu < 1$ TeV
\cite{EllisLH}, where only the upper mass bound gives cosmologically
relevant CDM. Instead of the neutralino, the axino, 
the supersymmetric partner of the axion, could be the lightest supersymmetric
particle. Axinos ($m_{\tilde{a}} < 300$ eV) would be a non-thermal relic, 
generated by the decay of neutralinos at $T \sim 10$~MeV \cite{axino}. 
Other non-thermal relics are WIMPzillas \cite{wimpzillas},
particles that are so heavy that they can never be in thermal equilibrium.
They would be generated at the end of cosmological inflation.  

\subsection{Neutralinos and other WIMPs}

Let me start the discussion with the lightest supersymmetric particle
\cite{Griest}, which is one of the most popular CDM candidates nowadays. 
In the minimal supersymmetric standard model (MSSM) \cite{MSSM}
the lightest supersymmetric particle most likely is the neutralino 
(when it is assumed to be stable). The neutralino is a mixture of the bino,
neutral wino, and the two neutral higgsinos. The lightest one is expected to
closely resemble a bino \cite{bino}. Assuming high slepton masses 
($>200$ GeV) and the unification of gaugino mass parameters $M_1$ and $M_2$ 
at the GUT scale, allows to derive constraints on the mass of the lightest 
neutralino from the LEP 2 experiments. The result is 
$m_{\tilde{\chi}} > 39$~GeV \cite{PDG}.  
Adding constraints from cosmology (CMB fits to a $\Lambda$CDM model) and 
from $b\to s\gamma$ decay, suggests that $100$ GeV $< m_{\tilde\chi} < 500$ 
GeV\footnote{The upper limit holds for $\tan\beta < 30 (45)$ for $\mu <0(>0)$
in the parameter space of the constrained MSSM.} \cite{Ellis}. 

It is essential to distinguish between
the chemical freeze-out and the kinetic decoupling of neutralinos: the
chemical freeze-out determines the amount of neutralinos today and occurs
when the annihilation rate of neutralinos drops below the Hubble rate,
$\Gamma_{\rm ann}/H < 1$. Soon after the neutralinos become non-relativistic, 
the rate for neutralino annihilation, $\Gamma_{\rm ann} = \langle v\sigma_{\rm
ann}\rangle n_{\tilde\chi}$, is suppressed by the Boltzmann factor in the 
number density of the neutralinos, $n_{\tilde\chi} \sim 
(m_{\tilde\chi} T)^{3/2}\exp(-m_{\tilde\chi}/T)$. The freeze-out temperature of
the neutralino \cite{Griest} is $T_{\rm cd} \sim m_{\tilde\chi} /20 > 5$ GeV, 
and
neutralinos are chemically decoupled at the QCD transition. We do not discuss
the physics of neuralino freeze-out in any detail here. Besides the 
annihilation of neutralinos, it is important to take into account 
coannihilation channels (e.g. a neutralino and a sfermion 
could `annihilate' into ordinary particles). If the masses of the lightest 
neutralino and other sparticles are similar, coannihilation plays an important
role. Also contributions from poles and thresholds have to be treated with 
care. For details we refer the reader to refs.~\cite{Ellis,ann}.

\begin{figure}
\label{fig12} 
\centerline{\includegraphics[width=0.65\textwidth]{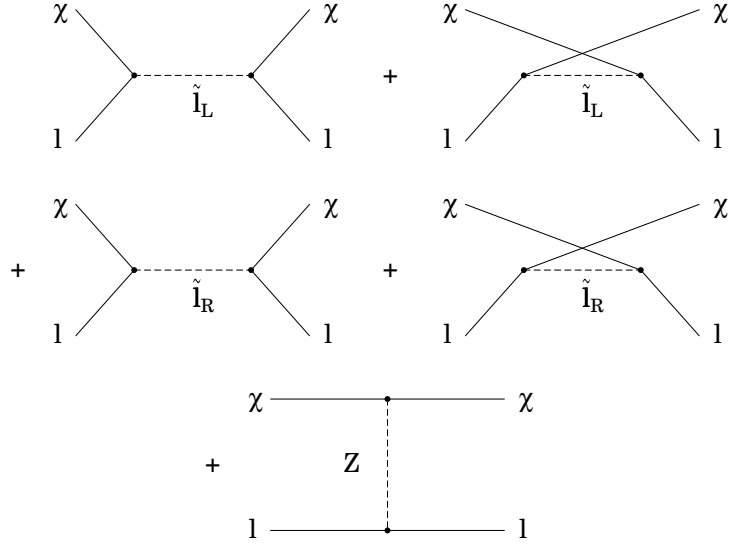}}
\caption{Feynman diagrams contributing to elastic neutralino--lepton
scattering; $\chi$ is the neutralino, $l$ is a lepton, and 
$\tilde{l}_{L,R}$ denote the corresponding left- and right-handed sleptons.} 
\end{figure}

Kinetic decoupling, in contrast, is determined by the elastic scattering
between neutralinos and the dominant radiation fluid. The interaction
rate for elastic scattering is $\Gamma_{\rm el} =\tau_{\rm coll}^{-1}= \langle
v\sigma_{\rm el} \rangle n$, where $n \sim T^3$ is the number density of
relativistic particles, e.g. electrons or neutrinos. The relevant diagrams for
elastic neutralino lepton scattering are shown in figure \ref{fig12}. 
Other contributions have been demonstrated to be of subleading order 
\cite{Chen}. 
If the neutralino is a pure gaugino, there is no contribution from $Z$ 
exchange. An order of magnitude estimate shows that 
$\langle\sigma_{\rm el}v\rangle 
\sim G_{\rm F}^2 T^2$ \cite{Griest88,HSS}, similar to elastic 
neutron--neutrino scattering \cite{Schramm}.   

One has to distinguish between the regime of perfect
kinetic coupling, i.e. neutralinos tightly coupled to the radiation fluid, an
intermediate regime where the neutralinos scatter
elastically but the number of collisions is not sufficient to drag them along
dissipationless, and the regime of free streaming, $\Gamma_{\rm el}/H<1$,
which is roughly at $T\le 1$ MeV, since the neutralino interacts weakly.

\begin{figure}[t]
\centerline{\includegraphics[width=0.65\textwidth]{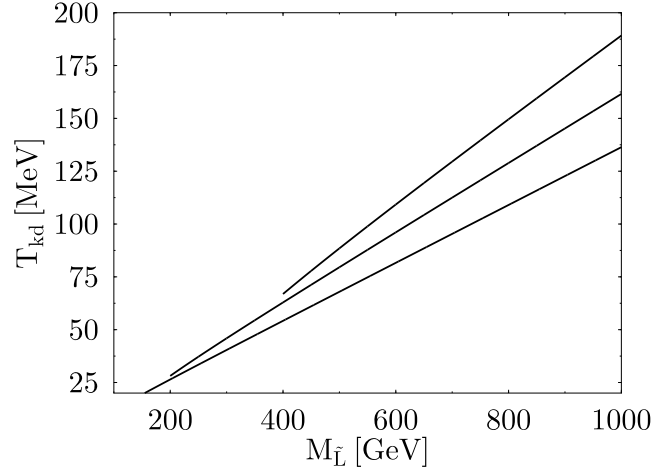}}
\caption{Kinetic decoupling 
as a function of the universal sfermion (slepton) mass
for bino masses $m_{\tilde{\chi}}\in\{50,100,200\}$ GeV
(from bottom to top) \protect\cite{HSS}.\label{fig13}}
\end{figure}

Let us estimate the regime where the neutralinos belong to the perfect
radiation fluid. Perfectness of a (dissipationless) fluid refers to an
external time scale, here $\omega^{-1}$ of an acoustic oscillation or the 
Hubble time. A fluid behaves as a perfect fluid if $\omega \tau_{\rm relax}
\ll 1$, i.e. the external time scale is larger than the relevant relaxation
time, the fluid is continually in local thermal equilibrium, and no entropy
production occurs. For the coupling of neutralinos to the radiation fluid the
relaxation time is given by $\tau_{\rm relax} = N \tau_{\rm coll}$, where $N$
is the number of collisions needed to completely change the momentum of the
neutralino due to collisions with the radiation fluid.
The momentum transfer at a collision of a lepton with the neutralino is of
order $p_{\rm l} \sim T$. The kinetic energy of the neutralino is given by
equipartition, $p_{\tilde\chi}^2/2 m_{\tilde\chi} \sim T$, hence its momentum 
is $p_{\tilde\chi} \sim \sqrt{m_{\tilde\chi} T}$. The fractional change of 
the neutralino
momentum from one collision at the QCD transition is $\delta p_{\tilde\chi}
/ p_{\tilde\chi} \sim \sqrt{T/m_{\tilde\chi}} \ll 1$. After $N$ collisions 
the total rms change of
momentum is $\left(\delta p_{\tilde\chi}/p_{\tilde\chi} \right)_{\rm rms} 
\sim \sqrt{N T/m_{\tilde\chi}}$. Local thermal equilibrium is obtained if 
the bulk motion of the
neutralinos is governed by the leptons, i.e. the fractional change of the
neutralino momentum is of order $1$. The number of collisions needed to
completely change direction is $N \sim m_{\tilde{\chi}}/T \approx 330$ for 
$m_{\tilde{\chi}} = 50$ GeV and $T=T_\star$. The collision time is given 
by the weak interactions rate $\Gamma_{\rm w} = \tau^{-1}_{\rm coll} 
\sim G_{\rm F}^2 T^5$, and the relaxation time is given by
\begin{equation}
\label{turnaround}
\tau_{\rm relax}= N \tau_{\rm coll} \sim N \times 10^{-7} t_{\rm H},
\end{equation}
with $N \sim m_{\tilde\chi}/T$. If the relaxation time is compared with the 
frequency of the acoustic oscillations, one finds that the condition for 
a perfect fluid, $\omega \tau_{\rm relax} \ll 1$, is satisfied, at the 
QCD transition, for scales
$\lambda> \lambda_{\tilde{\chi}-{\rm dec}}(T_\star) \approx 10^{-4} R_{\rm H}$ 
($m_{\tilde\chi} =
50$ GeV). Hence the neutralinos on these scales are part of the radiation
fluid at the QCD transition. Below this scale, the neutralinos cannot follow
the acoustic oscillations. On the other hand on the Hubble scale the perfect
kinetic coupling of neutralinos to the radiation fluid stops when the
required relaxation time becomes more than a Hubble time, $\tau_{\rm
relax}>t_{\rm H}$. This gives a temperature of $T_{\tilde{\chi}-{\rm dec}} 
\sim 10 $
MeV. The kinetic decoupling of neutralinos has been studied in detail in
\cite{HSS}. In figure \ref{fig13} the dependence of the kinetic decoupling
temperature on the neutralino mass (here a bino) and the slepton mass is 
shown. Down to the decoupling temperature neutralinos on the 
Hubble scale belong to the radiation fluid. 

The kinetic decoupling of other WIMPs, such as heavy neutrinos, also happens
way below $T_\star$. Therefore the neutralino or a heavy neutrino would 
be tightly coupled to the radiation fluid during the QCD transition. 

\subsection{Axions}

The axion has been introduced to solve the strong CP problem and it is a
popular CDM candidate \cite{rev}. Direct searches in the laboratory, 
constraints from astrophysics and from cosmology allow the axion mass to 
be $10^{-6}$ -- $3\times 10^{-3}$ eV\footnote{A stronger limit 
is obtained in \cite{Sakharov}, i.e. $m_{\rm a}  > 3.7 \times 10^{-4}$~eV.}. 
Axions could be the dominant matter 
today if their mass is small, i.e. $m_{\rm a} \sim 10^{-5}$~eV, which 
corresponds to a breaking of the Peccei--Quinn (PQ) symmetry at the 
scale $f_{\rm PQ} \sim 10^{12}$~GeV. These axions could be produced 
coherently by an initial misalignment of the axion field and by the decay 
of axionic strings (the latter is only possible if the reheating temperature 
after inflation lies above the PQ scale).  

The initially misaligned axion field starts to oscillate coherently when 
the axion mass has grown to $m_{\rm a}(T_1) \sim 3 H(T_1)$, where $T_1 
\sim 1$ GeV \cite{Turner86}. Below $T_1 \sim 1$ GeV the 
oscillating axion field evolves as CDM. This can be seen as follows:
the energy density of a scalar field is $\rho = \dot{\varphi}^2/2 + 
V(\varphi)$ and the pressure is $p = \dot{\varphi}^2/2 - V(\varphi)$. 
For time scales large with respect to the period of the coherent oscillations 
$\tau_{\rm a} \sim m_{\rm a}^{-1} \sim 10^{-6} t_{\rm H}$, the virial theorem
can be applied to give $\langle p_{\rm a} \rangle = 0$. Thus, a coherently
oscillating scalar field, like the axion field below $T_1$, can be modelled
by a pressureless fluid.  

The axion field is inhomogeneous on scales larger than $R_{\rm H}(T_1)$,
if, as is usually assumed, the reheating temperature after inflation
lies above $f_{\rm PQ}$. This leads to inhomogeneities with 
$\delta\rho_{\rm a}/\rho_{\rm a} = {\cal O}(1)$ in the axion density.
These inhomogeneities evolve into axion miniclusters
of typical mass $M_{\rm mc} \sim 10^{-12} M_\odot$ and typical radius
$R_{\rm mc} \sim 10^{8}$ m $\sim 0.1 R_\odot \sim 10^{-3}$ au \cite{Hogan}.
If the reheating temperature was below the
PQ scale, the axion field at $T_1$ was homogeneous on superhorizon scales,
apart from small, quantum-induced fluctuations of the axion field during
inflation. 

To summarize, the axion is a CDM candidate that is kinetically
decoupled from the radiation fluid at the QCD scale.
This is also the case for WIMPzillas and for primordial black holes.

\subsection{Primordial black holes}

A further possibility for CDM that decoupled kinetically long before the
QCD transition is primordial black holes (PBHs) \cite{CarrHawking,BHF} 
produced before the QCD transition and therefore with masses $M_{\rm BH} 
\ll 1 M_\odot$. In order to survive until today, PBHs should have $M_{\rm BH} 
> 10^{15}$g $\approx 10^{-18} M_\odot$. PBHs in the range from $10^{-18} 
M_\odot$ to $10^{-16} M_\odot$ would radiate too strongly to be compatible 
with $\gamma$-ray observations \cite{Carr,GL}. The production of PBHs arises 
naturally from hybrid inflationary scenarios \cite{inflBH,GL,Bullock,Kiefer}.

When these black holes are formed, they are comoving with the radiation fluid,
but their velocity is redshifted and soon becomes completely negligible. 
PBHs should not overclose the Universe, so that just a tiny fraction of energy 
in a comoving volume is allowed to collapse to black holes. This fraction 
is given by $f \sim a_{\rm PBH}/a_{\rm eq}$ , where $a_{\rm PBH}$ is the scale 
factor at PBH formation and $a_{\rm eq}$ at matter--radiation equality. 
Their density at the QCD epoch is 
\begin{equation}
\label{nPBH}
n_{\rm PBH}(T_\star) \sim 10^{-8} \left(150 \mbox{\ MeV}\over T_\star\right)
                          \left(T_{\rm PBH}\over T_\star\right)^2 
                          {1\over R_{\rm H}^3} 
\end{equation}
at most, where $T_{\rm PBH}$ is the temperature at which PBHs form with mass
\begin{equation}
\label{mPBH}
m_{\rm PBH} \sim \left( T_\star \over T_{\rm PBH} \right)^2 M_\odot \ .
\end{equation}
In every Hubble volume PBHs make up a pressureless gas if $n_{\rm PBH}(T_\star) 
\gg 1/ R_{\rm H}^3$, thus $T_{\rm PBH} > 10^4 T_\star \sim 1$~TeV. This 
corresponds to PBHs with $m_{\rm PBH} < 10^{-8} M_\odot$. PBHs smaller than 
this mass make a kinetically decoupled fluid at the QCD horizon scale.  

\section{Density fluctuations in CDM} 

\subsection{Kinetically decoupled CDM \label{initial}}

CDM is assumed to be a major component of the Universe today.
At the time of the QCD transition, however, the contribution of CDM
to the total energy density was negligible: 
$\epsilon_{\rm cdm}/\epsilon_{\rm rad} =
a/a_{\rm eq} \approx 10^{-8}$. Here, we consider a type of CDM which is
non-relativistic ($p \ll \epsilon$) at the QCD transition and which is only
coupled via gravity to the radiation fluid. Examples are axions or PBHs.
CDM moves in the external
gravitational potential provided by the dominant radiation fluid. During
a first-order QCD transition, the big amplifications of the density contrast 
in the radiation fluid $\delta_{\rm rad}$ (see section \ref{radpert}) 
leads to a big amplification in the gravitational potential. The CDM is
accelerated to higher velocities at the end of the transition. Therefore, we
also get peaks and dips in the cold dark matter fluid, as has been discussed
in \cite{SSW,SSW2}.

The initial conditions for CDM are obtained assuming 
adiabatic perturbations, i.e. the entropy per cold particle is unperturbed 
$\delta \left(s_{\rm rad}/n_{\rm cdm}\right) = 0$. Since $\epsilon_{\rm cdm} 
\propto 1/a^3$ and $\epsilon_{\rm rad} \propto 1/a^4$, the adiabatic initial 
conditions for $\delta_{\rm cdm}$ can be written $\delta_{\rm cdm}=
(3/4)\delta_{\rm rad}$. The initial fluid velocities are equal. 

\begin{figure}[t]
\centerline{\includegraphics[width = 0.65\textwidth]{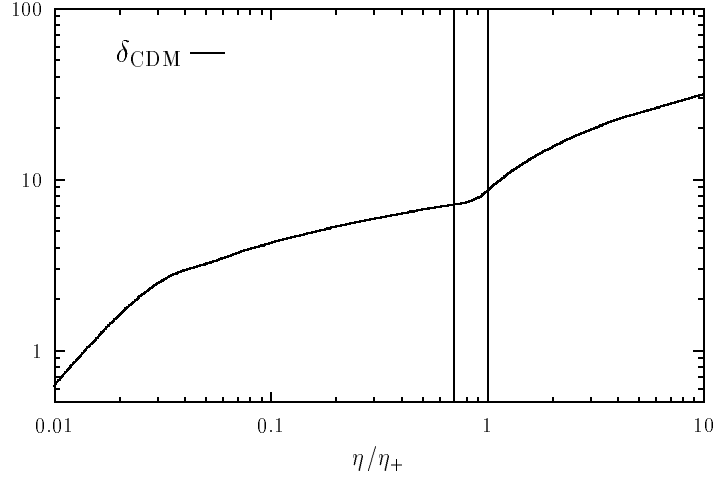}}
\caption{\label{fig14}
The time evolution of the CDM density contrast, $\delta_{\rm cdm}$ 
(from \protect\cite{SSW2}).
The major amplification of $\delta_{\rm cdm}$ is due to the higher
peculiar velocity at the end of the transition.}
\end{figure}

The subhorizon evolution of CDM in a purely radiation-dominated Universe is
just inertial motion, as can be seen from the Euler equation.
At leading order in $x=k/{\mathcal H}$, the
gravitational force can be neglected and the subhorizon evolution of the CDM
velocity is obtained:
\begin{equation}
\hat{\psi}_{\rm cdm} = C \frac{1}{x}.
\label{CDMinitial}
\end{equation}
The velocity of CDM in a radiation-dominated Universe just redshifts to
zero on subhorizon scales; $C$ is an integration constant
of order $A_{\rm in}$. The corresponding evolution
of the density contrast $\delta_{\rm cdm}$ follows from the continuity
equation,
\begin{equation}
\label{subCDM}
\delta_{\rm cdm} = C \ln{x} + D.
\end{equation}
This logarithmic growth of $\delta_{\rm cdm}$ \cite{Meszaros} can be seen in
fig.~\ref{fig14} before and after the transition. The shape of $\delta_{\rm
cdm}$ can also be seen in the transfer functions
of figs.~\ref{fig9} and \ref{fig10} on scales above the horizon 
scale $M_1$.

The major amplification effect comes from a 
higher velocity at the end of the transition,
which leads to
\begin{equation}
\label{growthCDM}
\delta_{\rm cdm}(\eta) =  \hat{\psi}^{\rm cdm}_+ k\eta_+
\ln\left({\eta\over\eta_+}\right) + \delta^{\rm cdm}_+.
\end{equation}
The amplification of the density contrast during the transition is negligible
with respect to the enhancement of the velocity. This velocity enhancement 
during
the transition leads to an additional logarithmic growth of $\delta_{\rm cdm}$
after the transition:
\begin{equation}
\delta_{\rm cdm}(\eta) =
\left[\frac34\left({\Delta\eta\over \eta_+}\right)^2
\hat{\psi}_-^{\rm rad}\right]
k\eta_+ \ln\left({\eta\over\eta_+}\right) + \delta^{\rm cdm}_+.
\end{equation}
This additional logarithmic growth of $\delta_{\rm cdm}$ is most
clearly seen in fig.~\ref{fig14}.

Let us compare the `QCD peaks' in the CDM spectrum with the CDM spectrum
without phase transition. In the limit $\epsilon_{\rm cdm}/\epsilon_{\rm rad}
\ll 1$, without transition, we find ($x = k/{\cal H}$)
in the subhorizon limit:
\begin{eqnarray}
\delta_{\rm cdm} &=& {3 A_{\rm in}\over 2}
\left[\ln\left({x\over\sqrt{3}}\right) +
\underbrace{\gamma_{\rm E} - \frac12}_{\approx 0.077} \right], 
\end{eqnarray}
which fixes the constants $C$ and $D$ in eqs.~(\ref
{CDMinitial}) and (\ref{subCDM}).

We are now able to calculate the enhancement factor
\begin{equation}
\label{enhancement}
{\cal E}(\eta) \equiv \left[\delta_{\rm transition}\over
\delta_{\rm no\ transition}\right]_{\rm cdm}(\eta) \ .
\end{equation}
The enhancement at the smallest scales $M_{\rm cdm} \sim 10^{-20} M_\odot$
to which our calculation applies and at some intermediate scale $M_{\rm cdm}
\sim 10^{-15} M_\odot$ reads
\begin{equation}
{\cal E}(\eta_{\rm eq}) \approx \left\{
\begin{array}{ll}
600 (20) & \mbox{bag\ model} \\
10  (2)  & \mbox{lattice\ fit}
\end{array} \right\} \quad
\mbox{for\ } 10^{-20} (10^{-15}) M_\odot \ .
\end{equation}
This shows that both models lead to large enhancements of the CDM density
contrast at equality for small enough scales. For the lattice fit, the
enhancement becomes important at scales below $M_{\rm cdm} \sim 10^{-15}
M_\odot$.

\subsection{CDM clumps \label{clumps}}

CDM in the form of axions or PBHs is not subject to damping as
the radiation fluid, thus the peaks in CDM will survive until structure
formation starts. The free streaming scale of CDM is way below our smallest
scales, because the initial velocity of axions or PBHs is completely
negligible. An overdensity of CDM in the form of axions or PBH (or any
other matter that is kinetically decoupled at the QCD epoch) decouples from
the cosmic expansion when its density contrast becomes non-linear,
$(\delta\rho/\rho)_R \sim 1$ (condition for turn around). It collapses and
virializes by violent gravitational relaxations and forms a clump of CDM. For
spherical collapse the final viral radius is half of the radius at turn around
\cite{Padmanabhan}. In the following, I present an updated version of the 
discussion in \cite{SSW2}.

We take a COBE \cite{COBE} normalized spectrum and allow for a tilt $|n-1| \le
0.2$. During the radiation-dominated regime, $(\delta \rho/\rho)_k$ for CDM
continues to grow logarithmically. In $(\delta\rho/\rho)^2_R$ another
logarithm comes from summing modes up to $k \sim 1/R$, where $R$ is the radius
of the window function. The enhancement factor ${\cal E}$ of CDM density
fluctuations during the QCD transition has been obtained in
eq.~(\ref{enhancement}). Putting these factors together we obtain the
amplitude of CDM perturbations of size $R$ at the time of equality:
\begin{equation}
\left(\delta\rho \over \rho\right)^{\rm cdm}_R (t_{\rm eq}) \approx
2 \times 10^{-4} \left( {k\over k_0} \right)^{n-1\over 2}
\left[\ln\left({k\over k_{\rm eq}}\right)\right]^{\frac32}
{\cal E}_k(t_{\rm eq})\ ,
\end{equation}
where $k_0$ denotes the comoving wave number of the mode crossing the horizon
today. In the following we assume $\omega_{\rm m} = 0.15$, and thus
$z_{\rm eq} \approx 3600$. For a CDM mass of $10^{-15}$ $(10^{-20}) M_\odot$ 
a tilt of $n - 1 = 0.2$ gives a factor $\approx 16$ $(23)$. The logarithms
contribute another factor $\approx 98$ $(144)$. The enhancement factor has been
calculated in section \ref{initial} to be ${\cal E}(t_{\rm eq}) 
\approx 2(10)$ for the
lattice QCD fit. Looking at $3\sigma$ ($= 3$ standard deviations) peaks, 
we find without tilt ($n-1=0$):
\begin{equation}
\left(\delta\rho \over \rho\right)^{3 \sigma, n=1}_R (t_{\rm eq}) \approx
0.1(0.9) \quad \mbox{for\ } M_{\rm clump} = 10^{-15} (10^{-20}) M_\odot \ .
\end{equation}
This implies that these clumps go non-linear at $z_{\rm nl}
\approx 360$ $(3200)$ and collapse to clumps of radius
$R_{\rm clump} \approx 23(0.05) R_\odot$. With tilt ($n-1=0.2$), we find
\begin{equation}
\left(\delta\rho \over \rho\right)^{3 \sigma, n=1.2}_R (t_{\rm eq}) \approx
2 (20) \quad \mbox{for\ } M_{\rm clump} = 10^{-15} (10^{-20}) M_\odot \ .
\end{equation}
These over-densities start to collapse even before $t_{\rm eq}$, because CDM
locally starts to dominate at $\sim 2(20) T_{\rm eq}$. This leads to clump
radii of $R_{\rm clump} \approx 1.2 (0.004) R_\odot$.

We conclude that the peaks in the CDM spectrum lead to clumps
of masses $10^{-20}$ -- $10^{-10} M_\odot$. Today, these clumps would
have a density contrast of $10^{10}$ -- $10^{17}$, where the lower value
corresponds to a $10^{-15} M_\odot$ clump from an untilted CDM spectrum, the
larger value is for a $10^{-20} M_\odot$ clump from a tilted CDM spectrum.
The evolution of these clumps in the late stages of structure formation
remains to be investigated (disruption, mergers, etc.).

For larger enhancement, e.g. if it should turn out that the latent heat
is larger than the value from present lattice QCD calculations, more compact
clumps are possible. These could be subject to femto-lensing 
\cite{femtolensing}.
With the values of the lattice fit \cite{SSW,SSW2}, the CDM clumps are not
compact enough to lie within the Einstein radius, which is  $R_{\rm E}
\sim 0.02 R_\odot$ for a $10^{-15} M_\odot$ clump.

The clumping of CDM changes the expected reaction rates for some dark 
matter searches, because some of the rates depend on the space-time 
position of the detector, star, or planet. Thus experiments looking 
for axion decay in strong magnetic fields \cite{Sikivie,rev} actually 
yield upper limits on the local axion interaction rate. It seems these 
experiments tell us that the Earth is not sitting in an axion cloud, 
if such clouds existed. 

\subsection{Kinetically coupled CDM}

Let me now turn to CDM candidates that belong to the radiation fluid 
at some point and decouple kinetically long after freeze-out. The most 
important example here is the neutralino, but the physical processes are 
the same for any WIMP that has been in thermal contact with the radiation
fluid at some point. I follow the discussion of \cite{HSS}, which are in
agreement with the recent findings in \cite{Berezinsky}.  
\smallskip 

\noindent Collisional damping 

During the process of kinetic decoupling the neutralinos acquire a finite
mean free path. Density inhomogeneities on scales of the diffusion
length are damped by the mechanism of collisional damping. It is convenient to
describe the CDM as an imperfect fluid. We have shown in \cite{HSS} that
the coefficient of heat conduction vanishes at the leading order in
$T/m_{\tilde\chi}$. Thus the dominant contribution to collisional damping comes
from bulk and shear viscosity. Since the energy of the CDM fluid can be 
transferred
to the radiation fluid, which acts here like an inner degree of freedom for the
CDM particles, the bulk viscosity does not vanish. Nevertheless, the radiation
fluid can be treated as a perfect fluid since $\epsilon_{\rm rad} \gg
\epsilon_{\rm cdm}$ at kinetic decoupling of the neutralinos. We calculated the
relevant coefficients of transport from the kinetic theory in \cite{HSS}. 
At linear order in the relaxation time the coefficients of shear and bulk 
viscosity become $\eta_{\rm visc} \approx n_{\tilde\chi} T \tau_{\rm relax}$ 
and $\zeta_{\rm visc} \approx (5/3)\eta_{\rm visc}$, respectively.

\begin{figure}[t]
\centerline{\includegraphics[width=0.6\textwidth]{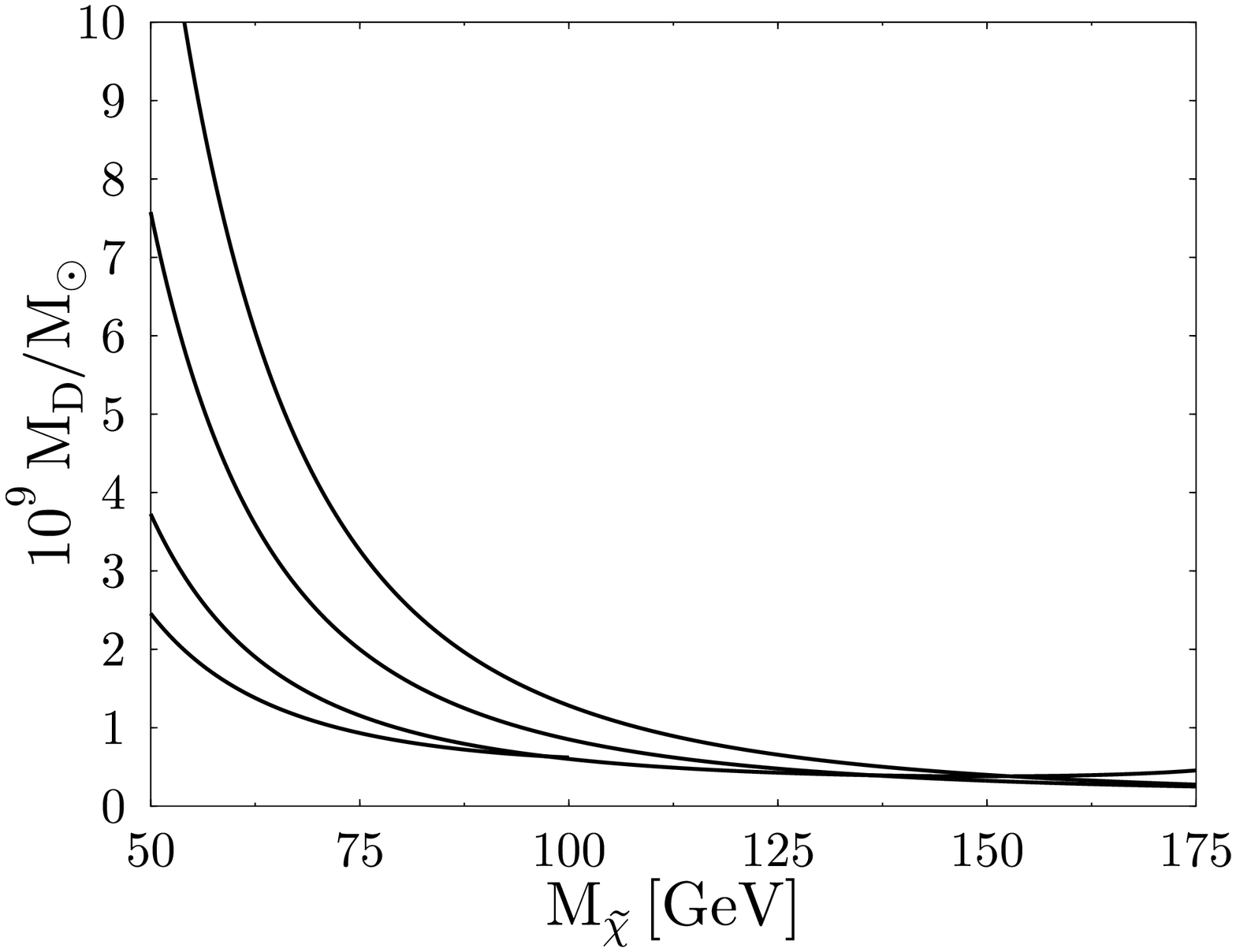}}
\caption{Acoustic damping scale as a function of the bino mass
for sfermion masses $M_{\widetilde{F}}\in\{150,200,300,400\}$ GeV
(from bottom to top) \cite{HSS}.\protect\footnote{Note that in the figure the 
CDM mass scale is larger that the CDM mass scale from the text by a 
factor of $8$, due to different definitions.}\label{fig15}}
\end{figure}

The density inhomogeneities in CDM are damped exponentially below the scale
$M_{\rm d}$ due to viscosity \cite{Weinberg,HSS}
\begin{equation}
\left({\delta\rho_{\tilde\chi} \over \rho_{\tilde\chi}}\right)_k \propto
\exp\left[-\frac 3 2 \int_0^{t_{\rm kd}}
\frac{T}{m_{\tilde\chi}}\; \tau_{\rm relax} k_{\rm ph}^2 {\rm d}t
\right] = \exp\left[ 
- \left(\frac{M_{\tilde{\chi}-{\rm dmp}}}{M}\right)^{2/3}\right].
\label{cd}
\end{equation}
In figure \ref{fig15} we plot the damping mass $M_{\tilde{\chi}-{\rm dmp}}$ 
as a function of the neutralino mass for various values of the slepton mass.
The damping (\ref{cd}) provides a small-scale cut-off in the primordial 
spectrum of density perturbations in neutralino CDM.
\smallskip 
 
\noindent Free streaming 

Once the temperature in the Universe drops below $T_{\rm kd}$ the rate of
elastic scatterings is not high enough to keep the neutralinos in thermal
equilibrium with the radiation fluid. The neutralinos enter the regime of
free streaming. This process continues to smear out inhomogeneities, since
the individual neutralinos do not move coherently. From kinetic theory one can
show that the damping due to free streaming goes as
\begin{equation}
\left({\delta\rho_{\tilde\chi} \over \rho_{\tilde\chi}}\right)_k \propto
\exp\left[- \frac{T_{\rm kd}}{2 m_{\tilde\chi}} \left(\frac{k_{\rm ph}}{H}
\right)^2_{T=T_{\rm kd}} \ln^2 \left(\frac{a}{a_{\rm kd}}\right)\right] =
\exp\left[ - \left(\frac{M_{\rm fs}(a)}{M}\right)^{2/3}\right].
\label{fs}
\end{equation}
The mass scale of damping from free streaming $M_{\rm fs}$ is written as
a function of the cosmic scale factor $a$. In the radiation-dominated 
Universe, the damping scale grows logarithmically with the scale factor. 
This calculation agrees with the estimate of the free streaming scale 
from the free streaming length as presented in \cite{HSS} up to a numerical 
factor $(2\pi/\sqrt{6})^3 \approx 17$. (We previously underestimated the 
free streaming mass by that factor.)
The ratio
\begin{equation}
\label{mfs}
\frac{M_{\rm fs}}{M_{\tilde{\chi}-{\rm dmp}}} = 
\left[\sqrt{\frac 5 3}\ln \frac{a}{a_{\rm kd}}
\right]^3
\end{equation}
exceeds unity for $a > 2.2 a_{\rm kd}$. Free streaming thus starts to dominate
the damping from collisional damping once the Universe has doubled its size
after kinetic decoupling. It is interesting to evaluate (\ref{mfs}) at the
time of matter--radiation equality, since this is the moment when CDM density
perturbations start to grow linearly with the expansion. For a kinetic
decoupling temperature $T_{\rm kd} = 40$ MeV and for $\omega_{\rm m} 
= 0.15$ we find $M_{\rm fs}(a_{\rm eq})/M_{\tilde{\chi}-{\rm dmp}} 
\approx 1.3 \times 10^4$, 
thus $M_{\rm fs}(a_{\rm eq}) \approx 8 \times 10^{-7} M_\odot$ for 
$m_{\tilde\chi} = 150$ GeV. 

Typically the free streaming mass at the time of equality is of the order
of $10^{-6}$--$10^{-5} M_\odot$, which is in striking contrast to claims
in the literature (see e.g.~\cite{Gurevich}) that the minimal mass for the
first objects would be $\sim 10^{-13} (150 \mbox{\ GeV}/m_{\tilde\chi})^3
M_\odot$. The huge difference with our result comes mainly from the false
assumption that kinetic decoupling occurs simultaneously with chemical
decoupling. 

To summarize, collisional damping and free streaming smear out
any primordial density inhomogeneities in neutralino CDM below
$\sim 10^{-6} M_\odot$. This implies that there is a peak (subhorizon CDM
density perturbations grow logarithmically during the radiation epoch) in the
power spectrum close to the cut-off and therefore we have found the minimal
mass for the very first objects, if CDM is made of neutralinos. This result
does not depend in a strong way on the parameters of the supersymmetric
standard model. Looking at $N \sigma$ overdense neutralino 
regions, we find that they go nonlinear at $z_{\rm nl} \approx 36 N$ for a 
flat spectrum of primordial fluctuations.  

According to the picture of the hierarchical formation of structures, 
these very
small first objects are supposed to merge and to form larger objects,
eventually galaxies and larger structures. It is unclear whether some of
the very first objects have a chance to survive. CDM simulations show 
structures
on all scales, down to the resolution of the simulation \cite{K,Moore}.
However, the dynamic range is not sufficient to deal with the first CDM
objects, so the fate of the first CDM clouds is an open issue. A cloudy
distribution of neutralino CDM in the galaxy would have important 
implications for direct and indirect searches for dark matter. 
A first attempt to study these consequences in detail has been made recently
\cite{Berezinsky}. It is estimated that a small fraction of these neutralino 
clouds survives tidal disruption and that these are probably enough to 
dominate a neutralino annihilation signal in our galaxy. 

\section{Summary and concluding remarks}

\begin{figure}[t]
\centerline{\includegraphics[width=0.8\textwidth]{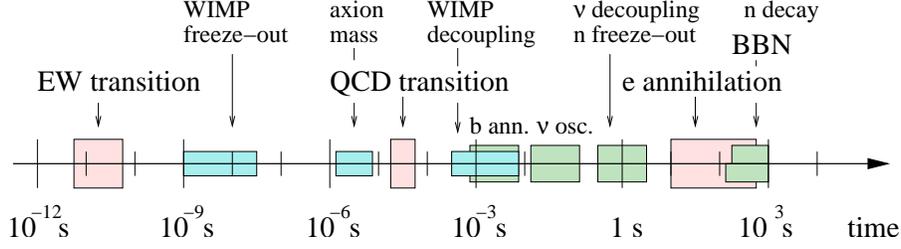}}
\caption{History of the first second of the Universe and the epoch of BBN. 
A sequence of thermodynamic transitions takes place: the electroweak 
transition ($\sim 10$ ps) is followed by the QCD transition ($\sim 10$ $\mu$s)
and by $e^\pm$~annihilation ($\sim 100$ s).
With respect to particles of the standard model, several important 
processes occur: baryons and antibaryons annihilate ($\sim 1$ ms), 
neutrino oscillations set in at $\sim 0.1$ s, neutrinos decouple and at the 
same time the neutron-to-proton ratio freezes out ($\sim 1$ s). During the 
epoch of BBN ($\sim 3$ min) this ratio changes slightly due to neutron decay 
($\tau = 886$ s). Regarding the cold dark matter candidates, WIMPs freeze out 
at $\sim 10$ ns and decouple kinetically at $\sim 1$ ms. The axion mass 
is switched on close to the QCD transition at $\sim 1$ $\mu$s.   
\label{fig16}}
\end{figure}

The general conclusion from this work is that many important cosmological 
issues are linked to the first second of the Universe, especially the 
nature of dark matter and the formation of nucleons.
In figure \ref{fig16} a summary of the history of that epoch is displayed. 

It might be that no relics from that epoch survive and it will 
therefore be hard or even impossible to probe the first second directly. 
Nevertheless, it is mandatory to study it in detail, since this is essential 
to answer the following questions: 
(i) Why are the initial conditions for the BBN (from observations)
very close to homogeneous? (ii) What is the small-scale structure of CDM 
and what is the implication of the CDM small-scale structure for
dark matter searches? (iii) Can we exclude that CDM is baryonic? (Are quark 
nuggets or other QCD-relics excluded)? 

Let me finally stress that this review represents my personal point of view 
and, there are certainly more issues than discussed here that are relevant 
for a complete understanding of the early Universe. 

\begin{acknowledgement}
I would like to thank S. Hofmann, J. Ignatius, J. Martin, C. Schmid and 
P. Widerin for precious and enjoyable collaborations and I am grateful to 
Z. Fodor, M. Laine, S. Kraml, D. Pav\'{o}n, M. Pl\"umacher, A. Rebhan, 
L. Roszkowski, A. S. Sakharov, S. Sanyal, B. Tom\`a\v{s}ik and X. Zhang 
for very useful comments and discussions. I thank S. Vascotto for 
proofreading and suggestions concerning the style. 
\end{acknowledgement}


\begin{table}[hb]
\begin{tabular}{ll}
\hline 
$a$ & scale factor \\
$B$ & depending on context, baryon number or bag constant \\
$c_s$ & speed of sound \\
$d$ & distance \\
$f$ & depending on context, frequency or volume fraction \\
$g$ & effective number of relativistic helicity degrees of freedom \\
$H$ & Hubble rate \\
$h$ & depending on context, $\equiv H_0/(100$ km$/$s$/$Mpc$)$ or amplitude 
      of gravitational waves \\
$k$ & comoving wave number \\
$L$ & lepton number \\
$l$ & latent heat density \\
$M$ & masses of macroscopic objects (with some noted exceptions) \\
$m$ & particle masses \\
$N_f$ & number of quark flavours \\
$n$ & number density \\
$p$ & pressure \\
$R$ & radius \\
$s$ & entropy density \\
$T$ & temperature \\
$T_*$ & transition temperature \\
$t$ & cosmic time \\
$v$ & velocity \\
$z$ & redshift \\
$\Gamma$ & interaction rate \\
$\epsilon$ & energy density \\
$\zeta_{\rm visc}$ & bulk viscosity \\
$\eta$ & depending on context, $\equiv n_{\rm B}/n_{\gamma}$ or conformal time\\
$\eta_{\rm visc}$ & shear viscosity \\
$\lambda$ & wave length \\
$\mu$ & chemical potential \\
$\rho$ & mass density \\
$\sigma$ & depending on context, surface tension or cross section \\
$\tau$ & time scale \\
$\Omega$ & fractional energy density \\
$\omega$ & $\equiv \Omega h^2$ \\
\hline
\end{tabular}
\caption{List of symbols.}
\begin{tabular}{llll}
\hline 
$0$ & today          & eq  & matter--radiation equality \\
ann & annihilation   & fs  & free streaming \\
b   & baryons        & gw  & gravitational waves \\
c   & critical       & H  & typical scale set by Hubble expansion \\
cd  & chemical decoupling & kd  & kinetic decoupling \\
cdm & cold dark matter & m  & matter $=$ baryons $+$ CDM \\
coll & collision     & mfp & mean free path \\
dec & decoupling     & nuc & nucleation \\
defl & deflagration  & ph  & physical \\
diff & diffusion     & rad & radiation \\
dmp & damping        & rms & root mean square \\
el  & elastic        & sc  & supercooling \\
\hline
\end{tabular}
\caption{Meaning of suffixes.}
\end{table}

\end{document}